\documentclass[twocolumn]{aastex62}
\usepackage{longtable}
\usepackage{comment}

\graphicspath{{./}{figures/}}

\submitjournal{AAS Journals}

\shorttitle{HD 1397b}
\shortauthors{Brahm et al. 2018}

\begin{document}

\title{HD~1397b: a transiting warm giant planet orbiting a V = 7.8 mag sub-giant star discovered by TESS}

\newcommand{\feh}{\ensuremath{{\rm [Fe/H]}}}
\newcommand{\teff}{\ensuremath{T_{\rm eff}}}
\newcommand{\teq}{\ensuremath{T_{\rm eq}}}
\newcommand{\logg}{\ensuremath{\log{g}}}
\newcommand{\zaspe}{\texttt{ZASPE}}
\newcommand{\ceres}{\texttt{CERES}}
\newcommand{\tess}{\textit{TESS}}
\newcommand{\vsini}{\ensuremath{v \sin{i}}}
\newcommand{\kms}{\ensuremath{{\rm km\,s^{-1}}}}
\newcommand{\mjup}{\ensuremath{{\rm M_{J}}}}
\newcommand{\mearth}{\ensuremath{{\rm M}_{\oplus}}}
\newcommand{\mpl}{\ensuremath{{\rm M_P}}}
\newcommand{\rjup}{\ensuremath{{\rm R_J}}}
\newcommand{\rpl}{\ensuremath{{\rm R_P}}}
\newcommand{\rstar}{\ensuremath{{\rm R}_{\star}}}
\newcommand{\mstar}{\ensuremath{{\rm M}_{\star}}}
\newcommand{\lstar}{\ensuremath{{\rm L}_{\star}}}
\newcommand{\rsun}{\ensuremath{{\rm R}_{\odot}}}
\newcommand{\msun}{\ensuremath{{\rm M}_{\odot}}}
\newcommand{\lsun}{\ensuremath{{\rm L}_{\odot}}}

\newcommand{\mptess}{\ensuremath{0.335_{-0.018}^{+0.018} }}
\newcommand{\rptess}{\ensuremath{1.021_{-0.014}^{+0.015} }}
\newcommand{\mst}{\ensuremath{1.284_{-0.016}^{+0.020} }}
\newcommand{\rst}{\ensuremath{2.314_{-0.042}^{+0.049} }}
\newcommand{\age}{\ensuremath{4.7 \pm 0.2 }}
\newcommand{\teqv}{\ensuremath{1213 \pm 17 }}

\newcommand{\per}{\ensuremath{11.53508 \pm 0.00057 }}
\newcommand{\sma}{\ensuremath{0.10866_{-0.00045}^{+0.00037} }}
\newcommand{\rhost}{\ensuremath{0.147 \pm 0.004 }}

\newcommand{\TC}{\ensuremath{2458332.08169 \pm 0.00046}}
\newcommand{\imp}{\ensuremath{0.16^{+0.14}_{-0.11}}}
\newcommand{\inc}{\ensuremath{89.1 $\pm$ 0.8}}
\newcommand{\ldca}{\ensuremath{0.146^{+0.036}_{-0.024}}}
\newcommand{\ldcb}{\ensuremath{0.85^{+0.10}_{-0.15}}}
\newcommand{\Krv}{\ensuremath{26.1^{+1.3}_{-1.3}}}
\newcommand{\ecc}{\ensuremath{0.210 \pm 0.038}}
\newcommand{\peri}{\ensuremath{0.19 \pm 0.20 }}
\newcommand{\rprs}{\ensuremath{0.04536^{+0.00035}_{-0.00026}}}
\newcommand{\ars}{\ensuremath{9.96^{+0.47}_{-0.47}}}
\newcommand{\rvslope}{\ensuremath{0.38^{+0.07}_{-0.06}}}
\newcommand{\sesinw}{\ensuremath{0.096^{+0.086}_{-0.091}}}
\newcommand{\secosw}{\ensuremath{0.435^{+0.029}_{-0.033}}}
\newcommand{\sigmatess}{\ensuremath{6^{+16}_{-4}}}
\newcommand{\muferos}{\ensuremath{30765.3^{+5.2}_{-4.3}}}
\newcommand{\mufideos}{\ensuremath{30695.3^{+5.2}_{-4.7}}}
\newcommand{\muharps}{\ensuremath{30804.4^{+5.1}_{-4.4}}}

\newcommand{\perc}{\ensuremath{18.24 \pm 0.44}}
\newcommand{\TCc}{\ensuremath{2458323.67355 \pm 1.8}}
\newcommand{\Krvc}{\ensuremath{7.5 \pm 1.4}}

\newcommand{\teffv}{\ensuremath{5479 \pm 50}}
\newcommand{\loggv}{\ensuremath{3.816 \pm 0.006}}
\newcommand{\fehv}{\ensuremath{+0.04 \pm 0.04}}
\newcommand{\vsiniv}{\ensuremath{4.0 \pm 0.3}}

\newcommand{\plname}{HD~1397b}
\newcommand{\stname}{HD~1397}
\newcommand{\rhopl}{\ensuremath{{\rm \rho_P}}}
\newcommand{\rhopkep}{\ensuremath{1.154 \pm 0.045 }}
\newcommand{\gccm}{\ensuremath{\mathrm{g}\,\mathrm{cm}^{-3}}}

\correspondingauthor{Rafael Brahm}
\email{rbrahm@astro.puc.cl}

\author[0000-0002-9158-7315]{Rafael Brahm}
\affiliation{Center of Astro-Engineering UC, Pontificia Universidad Cat\'olica de Chile, Av. Vicu\~{n}a Mackenna 4860, 7820436 Macul, Santiago, Chile}
\affiliation{Instituto de Astrof\'isica, Pontificia Universidad Cat\'olica de Chile, Av.\ Vicu\~na Mackenna 4860, Macul, Santiago, Chile}
\affiliation{Millennium Institute for Astrophysics, Chile}

\author[0000-0001-9513-1449]{N\'estor Espinoza}
\altaffiliation{Bernoulli fellow; Gruber fellow}
\affiliation{Max-Planck-Institut f\"ur Astronomie, K\"onigstuhl 17, Heidelberg 69117, Germany }

\author[0000-0002-5389-3944]{Andr\'es Jord\'an}
\affiliation{Instituto de Astrof\'isica, Pontificia Universidad Cat\'olica de Chile, Av.\ Vicu\~na Mackenna 4860, Macul, Santiago, Chile}
\affiliation{Millennium Institute for Astrophysics, Chile}

\author{Thomas Henning}
\affiliation{Max-Planck-Institut f\"ur Astronomie, K\"onigstuhl 17, Heidelberg 69117, Germany }

\author{Paula Sarkis}
\affiliation{Max-Planck-Institut f\"ur Astronomie, K\"onigstuhl 17, Heidelberg 69117, Germany }

\author{Mat\'ias I. Jones}
\affiliation{European Southern Observatory, Casilla 19001, Santiago, Chile}

\author{Mat\'ias R. D\'iaz}
\affiliation{Departamento de Astronom\'ia, Universidad de Chile, Camino El Observatorio 1515, Las Condes, Santiago, Chile}

\author{James S. Jenkins}
\affiliation{Departamento de Astronom\'ia, Universidad de Chile, Camino El Observatorio 1515, Las Condes, Santiago, Chile}

\author{Leonardo Vanzi}
\affiliation{Center of Astro-Engineering UC, Pontificia Universidad Cat\'olica de Chile, Av. Vicu\~{n}a Mackenna 4860, 7820436 Macul, Santiago, Chile}

\author{Abner Zapata}
\affiliation{Center of Astro-Engineering UC, Pontificia Universidad Cat\'olica de Chile, Av. Vicu\~{n}a Mackenna 4860, 7820436 Macul, Santiago, Chile}

\author{Cristobal Petrovich}
\affiliation{Canadian Institute for Theoretical Astrophysics, University
of Toronto, 60 St George Street, ON M5S 3H8, Canada}

\author{Diana Kossakowski}
\affiliation{Max-Planck-Institut f\"ur Astronomie, K\"onigstuhl 17, Heidelberg 69117, Germany }

\author{Markus Rabus}
\affiliation{Instituto de Astrof\'isica, Pontificia Universidad Cat\'olica de Chile, Av.\ Vicu\~na Mackenna 4860, Macul, Santiago, Chile}
\affiliation{Max-Planck-Institut f\"ur Astronomie, K\"onigstuhl 17, Heidelberg 69117, Germany }

\author{Pascal Torres}
\affiliation{Instituto de Astrof\'isica, Pontificia Universidad Cat\'olica de Chile, Av.\ Vicu\~na Mackenna 4860, Macul, Santiago, Chile}



\begin{abstract}

We report the discovery of a transiting planet first identified as a candidate 
in Sector 1 of the Transiting Exoplanet Survey Satellite (\tess), and then confirmed 
with precision radial velocities. \plname\ has a mass of \mpl = \mptess\ \mjup,
a radius of \rpl = \rptess\ \rjup, and orbits its bright host star ($V = 7.8$ mag) with an orbital period of \per\ d, on a moderately eccentric orbit ($e$ = \ecc). With a mass of 
\mstar = \mst\ \msun, a radius of \rstar = \rst\ \rsun, and an age of \age\ Gyr, the solar metallicity host star has already departed from the main sequence.
We find evidence in the radial velocity measurements for a long term acceleration, 
and a $P \approx 18$ d periodic signal that we attribute to rotational modulation 
by stellar activity.
The \stname\ system is among the brightest systems currently known to
host a transiting planet, which will make it possible to perform
detailed follow-up observations in order to characterize the
properties of giant planets orbiting evolved stars.

\end{abstract}



\section{Introduction} \label{sec:intro}
Throughout the past two decades, several ground-based, small-aperture,
wide-field photometric surveys \citep[e.g.][]{bakos:2004,pollacco:2006,pepper:2007,bakos:2013,talens:2017}
have efficiently detected and characterized the population of
short-period transiting giant planets orbiting bright stars across the
whole sky. These discoveries have triggered significant advances in
the study of the formation and evolution of planetary systems, and have
been targets of detailed follow-up observations to study their orbital
configurations \citep[e.g.][]{triaud:2010,zhou:2015,esposito:2017}
and atmospheric compositions \citep[e.g.][]{spake:2018,chen:2018,jensen:2018}.
Nonetheless, due to strong observational biases produced by the Earth's
atmosphere and the limited duty cycle of ground-based facilities,
there is still a region of parameter space of giant planets that is
vastly unexplored. Specifically, transiting systems of giant planets
having orbital periods longer than $\approx$10 d have scarcely been
detected from the ground \citep{kovacs:2010,wasp117,wasp130,brahm:2016:hs17}.
Additionally, due to the decrease in transit depth, we only have a 
handful of giant planets orbiting stars that have recently left the main sequence with radii larger than 2.2~\rsun \citep{hartman:2012,smith:2013,rabus:2016, bento:2018}.
The characterization of transiting giant planets having periods
longer than 10 d, and/or orbiting evolved stars is important for
understanding the processes that govern the formation, evolution
and fate of giant planets. The detailed study of the
distribution of eccentricities and spin-orbit angles of giant
planets orbiting
beyond 0.1 AU can be used to constrain migration theories \citep{dong:2014,petrovich:2016, anderson:2017}, while the characterization
of planets orbiting giant and sub-giant stars can be used to infer
the nature of the inflation mechanism of hot Jupiters \citep{lopez:2016}, characterizing
tidal interactions \citep[e.g.][]{villaver:2009}, and understanding the Lithium excess observed in some evolved stars \citep[e.g.][]{aguilera:2016}.
Very recently, the \textit{K2} mission \citep{howell:2014} started to contribute to the detection of
giant planets orbiting bright stars having $P>10$~d \citep[e.g.][]{shporer:2017,k2-232,k2-234,jordan:2018},
and giant planets orbiting evolved stars \citep[e.g.][]{grunblatt:2017, jones:2018}.
While not being its primary scientific driver, the \tess\ mission \citep{tess} is expected
to discover several hundreds giant planets orbiting bright stars (V$<$12 mag),
having periods longer than 10 d, and/or orbiting evolved stars \citep{barclay:2018}. Therefore,
it is expected that \tess\ will yield for the first time a statistically significant sample for these kind of objects that will be useful for constraining theories of planetary formation and evolution. Interestingly, the first confirmed giant planet from \tess\ orbits a slightly evolved star \citep{wang:2018}.

In this study we present the discovery of HD~1397b, the first giant planet discovered by the \tess\ mission with an orbital period longer than 10 days. In addition to its long period, the host star is  a bright ($V=7.8$) sub-giant star, making this a very interesting system for further study. 

\section{Observations} 
\label{sec:obs}

\subsection{TESS} 
\label{sec:tess}
Between 2018 July 25 and 2018 August 22, the \tess\ mission
observed \stname\ (TIC 394137592, TOI00120.01) with its
Camera 3, during the monitoring of the first \tess\ sector.
Observations were performed with a cadence of 2 minutes.
Photometric data of \stname\ were analyzed with the Science
Processing Operations Center (SPOC) pipeline, which is
a modified version of the pipeline used for the NASA Kepler
mission (Jenkins et al., in prep). This light curve was
released to the community on September 15 of 2018 as an
alert. The alert did not report warning flags that could
be associated with false positive
scenarios. In particular, there is no statistical difference
between transit depths, and there are no significant
centroid offsets of the PSF during the transits.

\begin{figure*}
\plotone{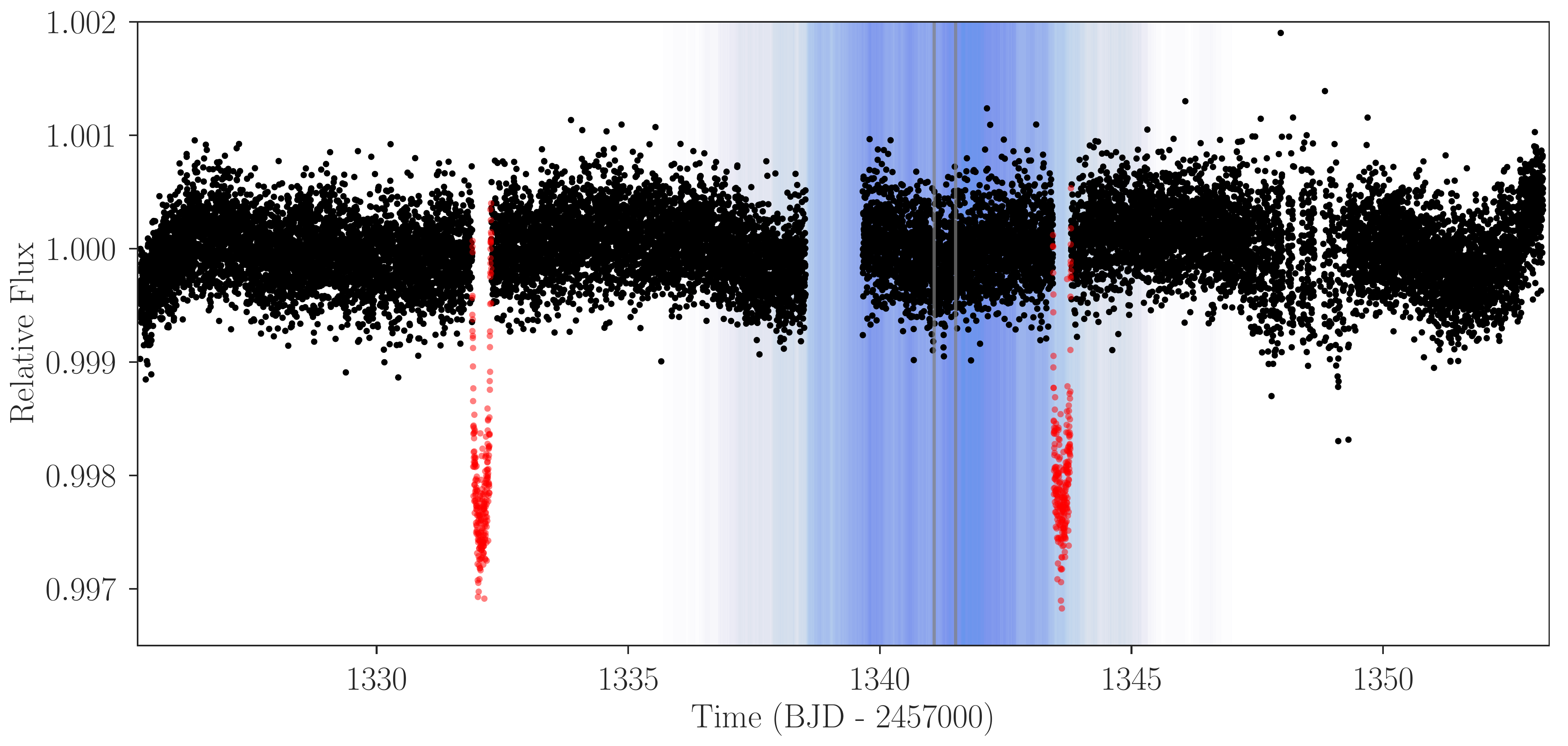}
\caption{\tess\ 2 minute cadence light curve for \stname. The two transits are plotted in red. The blue region represents the probability distribution for a transit produced by a planet having orbital parameters equal to those determined for the second radial velocity signal (see Section \ref{sec:glob})}
\label{fig:phot}
\end{figure*}

\begin{figure*}
\plotone{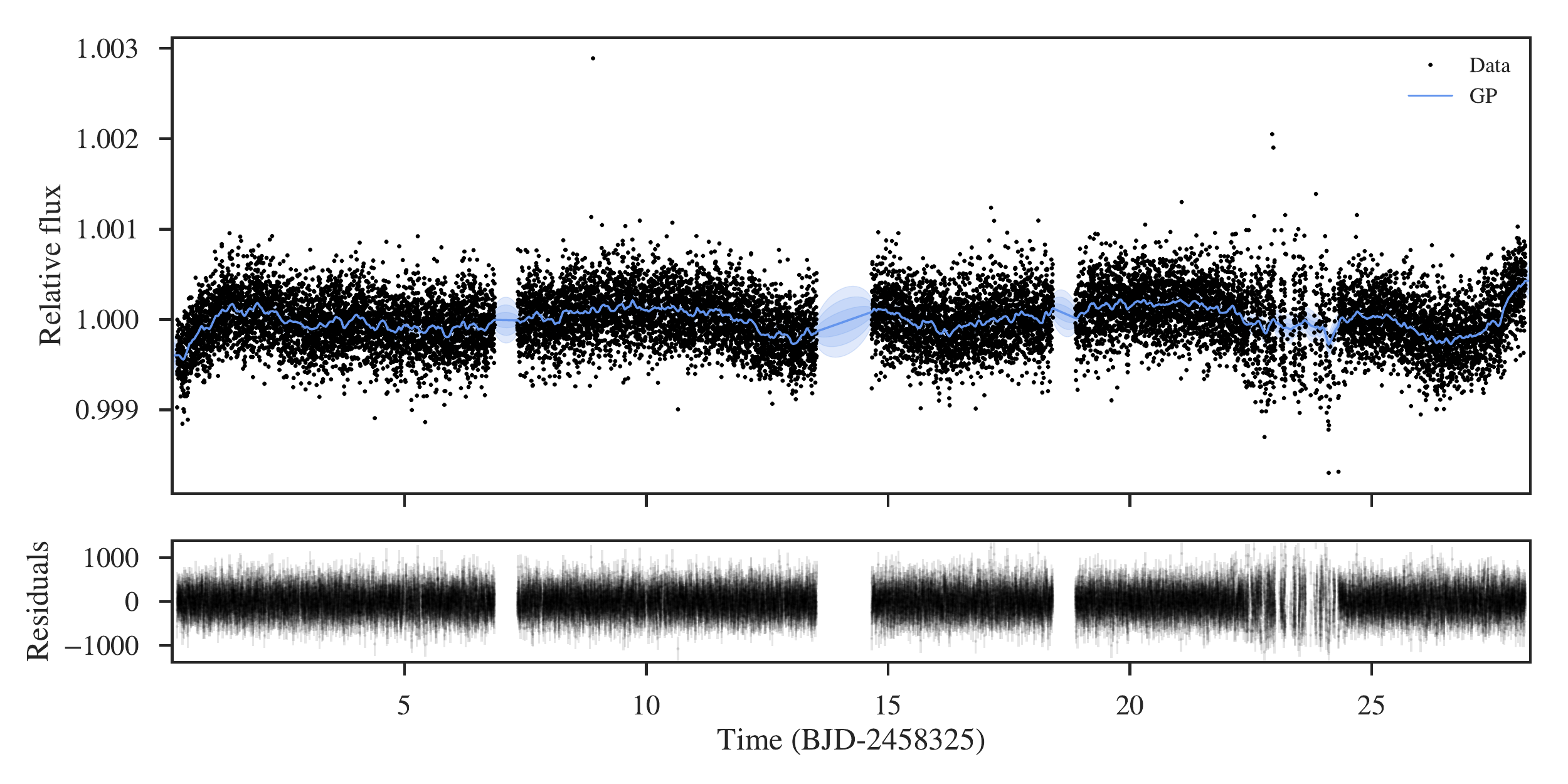}
\caption{\tess\ lightcurve with the transits masked out and a quasi-periodic kernel fitted to the data. This kernel was used to 
detrend the lightcurve and, especially, remove predicted long 
term trends in-transit.}
\label{fig:phot2}
\end{figure*}

The \tess\ light curve of \stname\ is presented in Figure \ref{fig:phot},
after removing flagged data points, and shows the presence of two clear, moderately deep ($\approx$2500 ppm) transits separated by $\approx$11.5 days. 
We masked out the in-transit points and ran a Gaussian process (GP) regression with the quasi-periodic 
kernel introduced in \citep{celerite} in the data, in order to both detrend 
the lightcurve and predict the in-transit trends, allowing for a photometric jitter term in 
this fit. The best-fit GP is shown in Figure \ref{fig:phot2}, and this GP was divided into our data in order to compute a detrended lightcurve, propagating the prediction errors into the 
photometric errors. As can be seen, the in-transit prediction shows a decreasing slope 
on the second transit, whereas in the first transit the 
GP prediction is relatively flat. 

\subsection{High resolution spectroscopy} 
\label{sec:spec}
We started the radial velocity follow-up of \stname\ a couple of hours
after the first alerts of \tess\ Sector 1 were made public. We obtained
two spectra with the HARPS spectrograph \citep{mayor:2003} installed at the ESO~3.6m telescope at La Silla Observatory on two consecutive nights. We adopted an exposure time of 300s and pointed the comparison fiber to the background sky. Additionally, we obtained 35 spectra with the FIDEOS spectrograph \citep{vanzi:2018} installed at the 1-m telescope on the same observatory.
FIDEOS observations were obtained on six consecutive nights. For these observations we adopted an exposure time of 600 s and the comparison fiber was used to trace the instrumental radial velocity drift by observing a ThAr lamp. The HARPS and FIDEOS data were processed with the \ceres\ package \citep{brahm:2017:ceres}, which automatically performs optimal spectral extraction from the raw images, wavelength calibration, instrumental drift correction, and the computation of the radial velocities and bisector spans. The radial velocities were computed with the cross-correlation technique by using a binary mask with a set of lines compatible with a G2-type star. These radial velocities were consistent with an absence of large variations that could have been produced by a stellar companion, and they hinted the presence of a relatively low amplitude signal ($\approx$ 30 m~s$^{-1}$) in phase with the photometric ephemeris. Additionally, no secondary peaks were identified in the cross-correlation function that could have indicated the presence of possible background diluted eclipsing binaries.

\begin{figure*}
\plotone{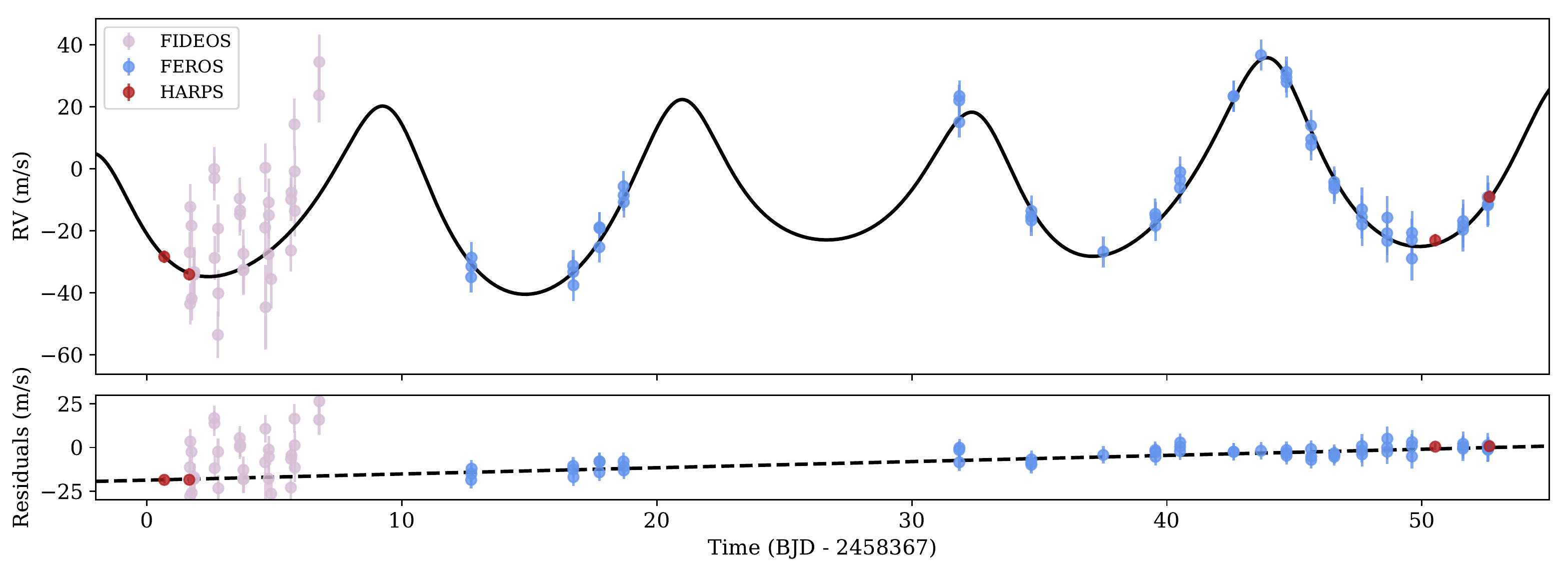}
\caption{The top panel presents the radial velocity (RV) curve for \stname\ obtained with HARPS (red), FIDEOS (pink), and FEROS (blue). The black line corresponds to the Keplerian model with the posterior parameters found in Section \ref{sec:glob}. The bottom panel shows the residuals without considering the radial velocity trend.\label{fig:rvstime}}
\end{figure*}

\begin{figure}
\plotone{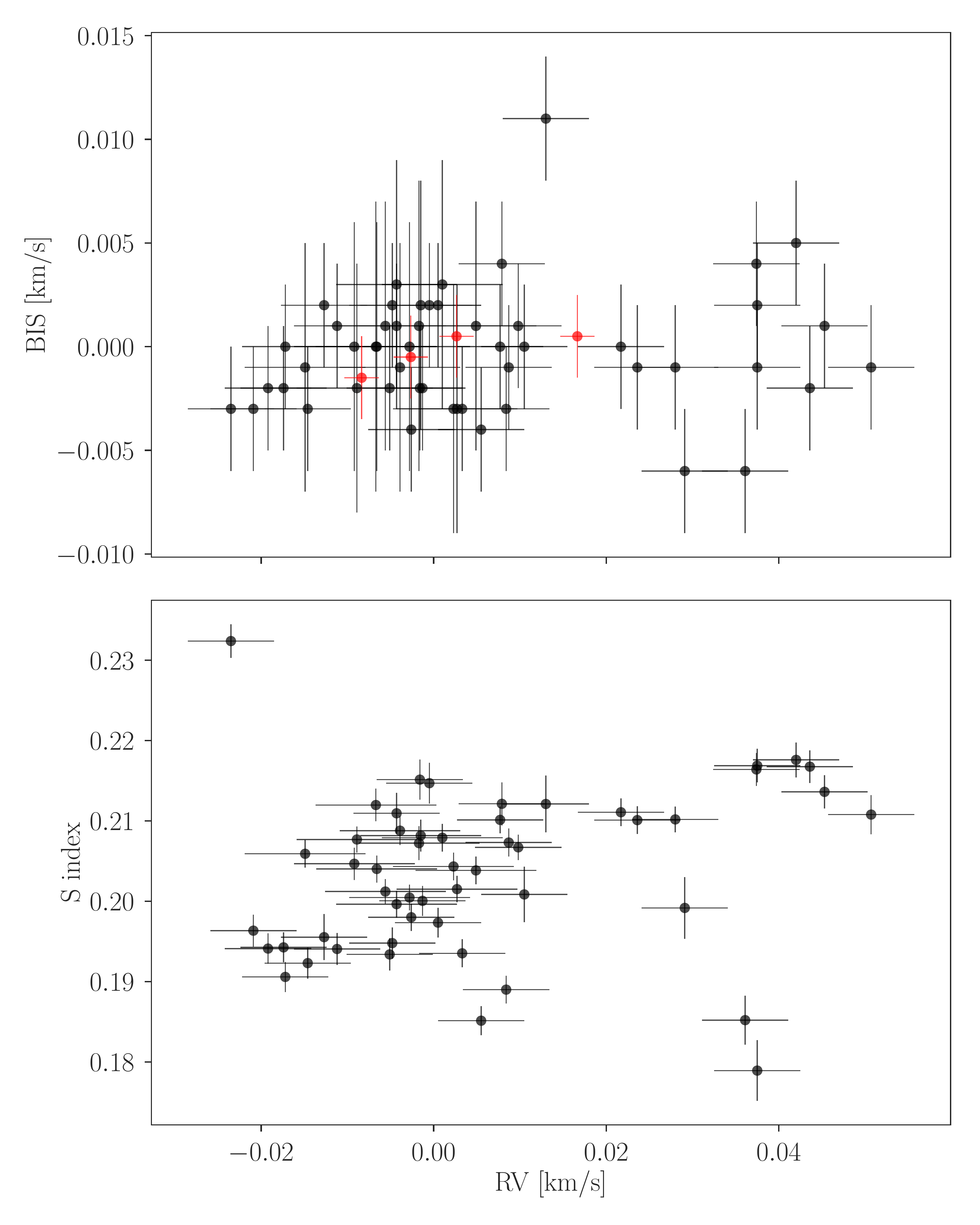}
\caption{Top panel: Radial velocity -- bisector span scatter plot for the FEROS (black) and HARPS (red) observations of \stname. No correlation is observed. Bottom panel: Radial velocity -- S-index scatter plot for the FEROS observations. While the correlation
value is larger in this case, it is still not significant at the 95\% confidence level.\label{fig:bis}}
\end{figure}

We then proceeded to perform intensive radial velocity monitoring of \stname\ with the FEROS spectrograph installed at the MPG~2.2m telescope at La Silla Observatory \citep{kaufer:99}. Two additional HARPS spectra were acquired
during this period. FEROS observations were performed with the simultaneous calibration mode and the adopted exposure time was of 300s. We obtained 52 spectra on time span of 40 nights. FEROS data were also processed with the \ceres\ package. In addition to the bisector span measurements, the S-index activity indicator was computed for each spectrum  as described in \citet{jenkins:2008,jones:2017}. \stname\ presents significant
and time variable chromospheric emission as gauged from the Ca\,{\sc ii} H \&  K lines of the FEROS and HARPS spectra.
The radial velocities, bisector spans, and S-indexes obtained with the three instruments are presented in Table~\ref{tab:rvs}. The velocities are  plotted as a function of time in Figure~\ref{fig:rvstime}.
The FEROS velocities confirmed the presence of the periodic signal initially hinted by FIDEOS, and also allowed the detection of a long term radial velocity trend. Additionally, no significant correlation was found between the radial velocity and bisector span measurements at the 95\% level of confidence (see Figure \ref{fig:bis}) A greater correlation was found between the S-index and the radial velocity measurements but it was still not significant at the 95\% level of confidence. The relation between the radial velocity, bisector span, and activity measurements are further explored in Section~\ref{sec:gls}.

\label{sec:chat}

\section{Analysis} 
\label{sec:anal}

\subsection{Properties of the host star} 
\label{sec:star}
We analyzed the co-added FEROS spectra to obtain the atmospheric
parameters of \stname. We used the \zaspe\ code \citep{brahm:2016:zaspe}, which
compares the observed spectrum to a grid of synthetic models generated
from the ATLAS9 model atmospheres \citep{atlas9}. The search for the optimal model
is performed only in the spectral zones that are most sensitive to changes in the atmospheric parameters. The errors in the derived parameters are
obtained from Monte Carlo simulations where the depth of the spectral
lines of the synthetic models are perturbed in order to properly account for the systematic model mismatch as the main source of the error budget. We obtained that \stname\ has an effective temperature of \teff = \teffv\ K, a surface gravity of \logg = \loggv\ dex, a metallicity of \feh = \fehv\ dex, and a projected rotational velocity of \vsini = \vsiniv\ m s$^{-1}$.

In order to estimate the radius of \stname, we followed the same procedure presented in \citet{k2-232}. We interpolated the stellar models presented in \citet{baraffe:2015} to generate a synthetic spectral energy distribution (SED) consistent with the atmospheric parameters found for \stname.
We then generated synthetic magnitudes from the SED by integrating it in different spectral zones, weighted by the corresponding
transmission functions of the passband filters presented in Table~\ref{tab:stprops}. 
These synthetic magnitudes were used to estimate the stellar radius (\rstar) and the extinction factor ($A_V$) by comparing them to the publicly reported magnitudes after converting them to absolute magnitudes using the distance obtained from the GAIA DR2 parallax measurement \citep{gaia,gaia:dr2}. 
We obtained the posterior distribution for \rstar\ and $A_V$ by using the \texttt{emcee} \texttt{Python} package \citep{emcee:2013}.
In order to consider the uncertainty in the atmospheric parameters, we  repeated this process for different values of \teff\ sampled from a Gaussian distribution with the parameters obtained with the \zaspe\ analysis. We find that \stname\ has a radius of \rstar\ = \rst\ \rsun, and an extinction factor consistent with $A_V=0$. 

Finally, we estimated the mass and evolutionary stage of \stname\ by comparing its effective temperature and radius to those predicted by stellar evolutionary models.
Specifically, we used the Yonsei-Yale evolutionary models \citep{yi:2001} which were interpolated to the metallicity derived for
\stname. We used the \texttt{emcee} package to compute the posterior distributions for the stellar mass and age. We find that \stname\ has a mass of \mstar = \mst\ \msun, and an age of \age\ Gyr. The stellar parameters obtained for \stname\ are presented in Table~\ref{tab:stprops} along with its observed properties.

\begin{deluxetable*}{lrc}[b!]
\tablecaption{Stellar properties of \stname\  \label{tab:stprops}}
\tablecolumns{3}
\tablewidth{0pt}
\tablehead{
\colhead{Parameter} &
\colhead{Value} &
\colhead{Reference} \\
}
\startdata
Names \dotfill   &    \stname  &   \\
 &  	TIC 394137592 & TESS \\
 &  	HIP 1419 & HIPPARCOS \\
 & 2MASS J00174714-6621323 & 2MASS  \\
 & TYC 8846-638-1 & TYCHO  \\
RA \dotfill (J2000) &  00h17m47.14s &  TESS\\
DEC \dotfill (J2000) & -66d21m32.35s &   TESS\\
pm$^{\rm RA}$ \hfill (mas yr$^{-1}$) & 64.76 $\pm$ 0.05& GAIA\\
pm$^{\rm DEC}$ \dotfill (mas yr$^{-1}$) & -5.06 $\pm$ 0.04 & GAIA\\
$\pi$ \dotfill (mas)& 12.54 $\pm$ 0.03 & GAIA \\ 
\hline
T \dotfill (mag) & 7.14 $\pm$ 0.03 & TESS\\
B  \dotfill (mag) & 8.47 $\pm$ 0.05 & APASS\\
V  \dotfill (mag) & 7.79 $\pm$ 0.03 & APASS\\
J  \dotfill (mag) & 6.44 $\pm$ 0.02 & 2MASS\\
H  \dotfill (mag) & 6.09 $\pm$ 0.05 & 2MASS\\
K$_s$  \dotfill (mag) &5.99 $\pm$ 0.02 & 2MASS\\
WISE1  \dotfill (mag) & 6.02 $\pm$ 0.09 & WISE\\
WISE2  \dotfill (mag) & 5.90 $\pm$ 0.04 & WISE\\
WISE3  \dotfill (mag) & 5.99 $\pm$ 0.02 & WISE\\
\hline
\teff  \dotfill (K) & \teffv & \texttt{zaspe}\\
\logg \dotfill (dex) & \loggv & \texttt{zaspe}\\
\feh \dotfill (dex) & \fehv & \texttt{zaspe}\\
\vsini \dotfill (km s$^{-1}$) & \vsiniv & \texttt{zaspe}\\
\mstar \dotfill (\msun) & \mst & YY + GAIA\\
\rstar \dotfill (\rsun) & \rst & GAIA + this work\\
L$_{\star}$ \dotfill (L$_{\odot}$) & 4.32 $\pm$ 0.19 & YY + GAIA\\
M$_{V}$ \dotfill (mag) & 3.30 $\pm$ 0.06 & YY + GAIA\\
Age \dotfill (Gyr) & \age & YY + GAIA\\
$\rho_\star$ \dotfill (g cm$^{-3}$) & \rhost  & YY + GAIA \\
\enddata
\end{deluxetable*}

\subsection{Radial Velocities} \label{sec:gls}
\begin{figure*}
\plotone{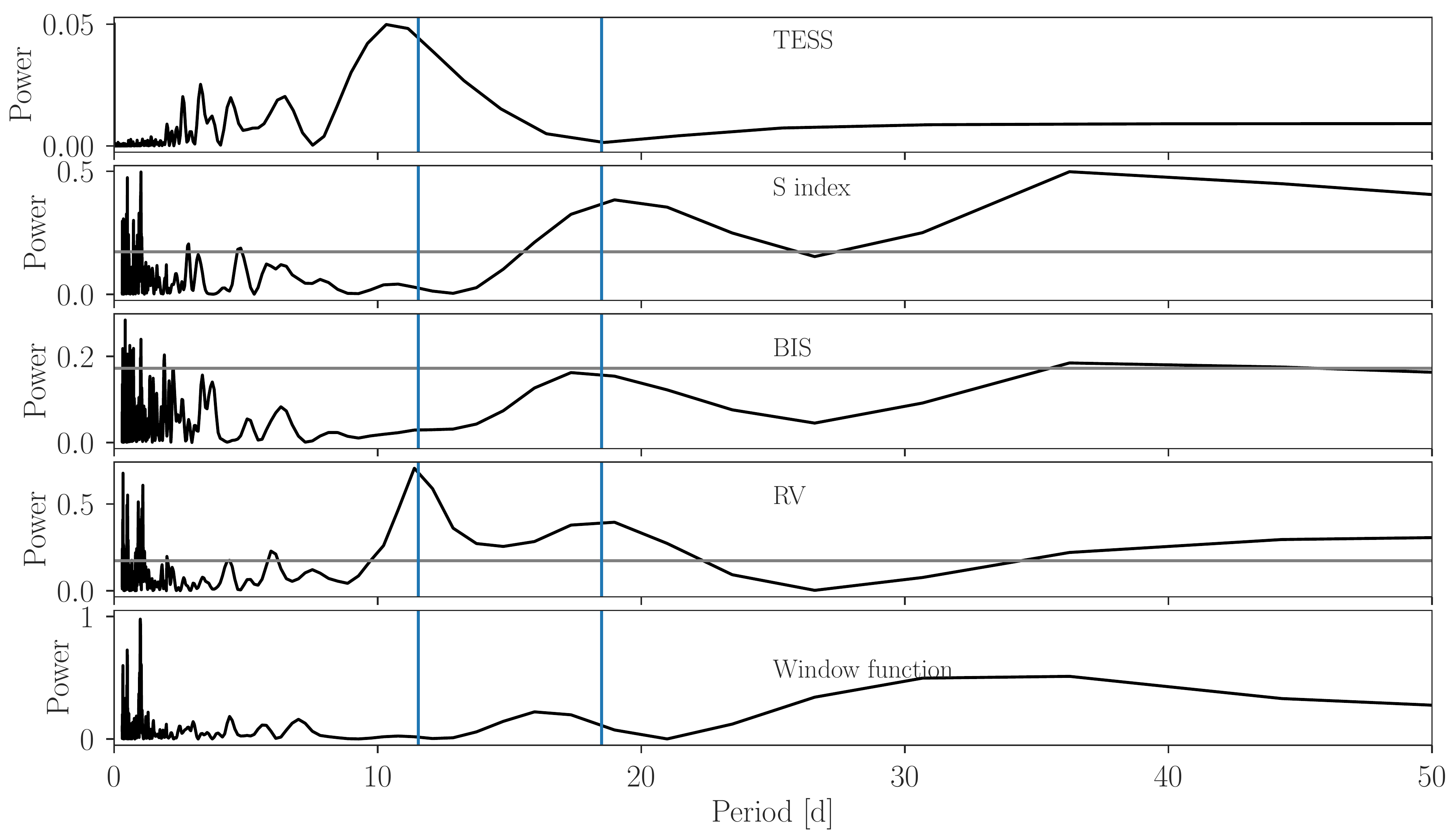}
\caption{GLS power spectra for the \tess\ photometry, S-index, bisector spans, radial velocities of \stname\, and window function (black lines from top to bottom). The gray lines represent the 1\% significance limits based on bootstrap simulations. The radial velocities present a primary peak at $\approx$11.5 days which is consistent with the orbital period of the planetary candidate obtained from the \tess\ photometry. There is a secondary peak in the periodogram of the radial velocities at $\approx$18.5 days, which can be associated with the rotational modulation of activity regions in the surface of the star.\label{fig:gls}}
the \end{figure*}

Motivated by the slight but not significant correlation between
the radial velocities and the activity indicators, and also
due to the presence of time correlated residuals in some
preliminary modeling of the radial velocities, we proceeded
to analyze all the available time series. We computed the
generalized Lomb-Scargle periodograms \citep[GLS, ][]{zechmeister:2009} for the non-detrended \tess\ photometry,
radial velocities, bisector span measurements, and S-index values.
For the \tess\ photometry we masked out the two transits. The
significance of the peaks in the periodograms was evaluated
by performing a bootstrap of each time series. The four
periodograms are plotted in Figure \ref{fig:gls}.
The bottom panel of Figure \ref{fig:gls} shows that
the radial velocity measurements are able by themselves to 
recover with high significance the 11.5 day periodic signal
of the planet candidate observed in the \tess\ transits.
Additionally, there are no peaks in the periodograms
of the bisector spans and S-index associated to that
particular period. Interestingly, the periodogram of the
\tess\ photometry presents a peak close to the
orbital period of the system. If the associated signal is
not an artifact of the SPOC pipeline, given the large
ratio between the radii of the star and the planet,
the source of this periodic variation could be
associated to changes on the star \citep[e.g. ellipsoidal variations,][]{welsh:2010}
rather than phase curve variations.

Figure \ref{fig:gls} shows also the presence of
a secondary wide significant peak at $\approx$18.5 days
in the radial velocity periodogram.
This periodic signal is also present with high significance in the
S-index time series, and marginally present in the bisector
span measurements. These signals can be associated to
the rotational modulation of active regions on the
stellar surface. Given the relatively short time
span of our spectroscopic observations (2 months),
the active regions on the star, such as spots,
faculae and/or plages could have remain stable producing
a coherent signal \citep{hussain:2002}. If the rotation
period of \stname\ is of $\approx$18.5 d, by using the
estimated radius of \rstar\ = \rst\ \rsun, we predict
a rotational velocity at the surface of the star of
$\approx$6.3 km s$^{-1}$, which if compared with the \vsini\ value
estimated from our spectroscopic analysis would imply a miss-alignment
angle of $\approx$60 degrees between the spin of the star
and the orbit of \plname.

\subsection{Global Modeling} \label{sec:glob}
The global modelling of the transit and radial velocities was 
performed with a new algorithm that will be detailed in another 
publication (Espinoza et al., in prep). In brief, the algorithm 
is similar to \texttt{exonailer} \citep{espinoza:2016:exo} with 
the main difference that instead of \texttt{emcee} \citep{emcee:2013}, 
the new algorithm uses MultiNest \citep{MultiNest} via the PyMultiNest package 
\citep{PyMultiNest} in order to both perform posterior sampling and model comparison 
directly using model evidences. The transits are modelled using \texttt{batman} \citep{kreidberg:2015}, 
and the radial velocities modelled using \texttt{radvel} \citep{fulton:2018}. 
For limb-darkening, we use a quadratic law with the uninformative 
sampling scheme of \cite{Kipping:LDs}. This enhanced algorithm in turn also has the possibility 
to include dilution factors, and variable mean fluxes for different photometric instruments, as 
well as the possibility of fitting multiplanetary systems for both transits and radial velocities 
simultaneously. Several types of trends as a function of time can also be modelled in the radial 
velocities including linear and quadratic trends. For these latter trends, the fitting parameters 
in our approach are the terms that accompany the linear and quadratic terms, along with a parameter 
$T_{rv,0}$, which is the time at which this linear trend crosses the line of zero radial velocity once 
the planetary signals and instrumental offsets of the radial-velocities are removed. This parametrization 
is made such that a convenient prior can be defined for this parameter. 

We considered several fits for our data, which showed not only an evident linear trend, but also an 
extra periodic signal that appeared in the residuals once we substracted a fit considering only the 
transiting exoplanet and said linear trend, hinting at a period of about 18 days, in line with the 
observed periodograms shown in Figure~\ref{fig:gls}. We thus fitted an 
extra keplerian signal to the radial-velocities in our joint analysis with a large prior around this 
period, which in turn gave a larger evidence for this model later ($\ln Z > 5$, i.e., a Bayes factor assuming both 
models are equally likely of $\sim 150$ in favor of the model containing the extra signal).
Interestingly, the posterior parameters for the secondary signal
imply that if it is due to another transiting planet, there is
a high probability that the transit would have occurred during the
the gap in the \tess\ light curve produced during the downlink of the data
from the satellite (Figure \ref{fig:phot}).
However, considering the signals observed in the periodograms
for other activity indicators (see Figure \ref{fig:gls}),
we consider it more likely that this signal is generated by stellar activity.
The parameters obtained from the global modeling 
are presented in Table~\ref{tab:plprops}.

By combining the results of the global analysis with the physical
parameters derived for the stellar host, we find that \plname\
has a Saturn-like mass of \mpl = \mptess\ \mjup, but a radius that
is more similar to that of Jupiter (\rpl = \rptess\~\rjup).
\plname\ has a semi-major axis of $a$=\sma\ AU, which coupled to
its orbital eccentricity results in a distance to
the star at periapsis of 0.08306 $\pm$ 0.00003~AU.
This orbital configuration also results in  an
averaged equilibrium temperature for \plname\ of \teq = \teqv\ K,
assuming zero albedo and full energy redistribution.
The derived parameters of \plname\ are listed in Table~\ref{tab:plprops}.

\begin{figure}
\plotone{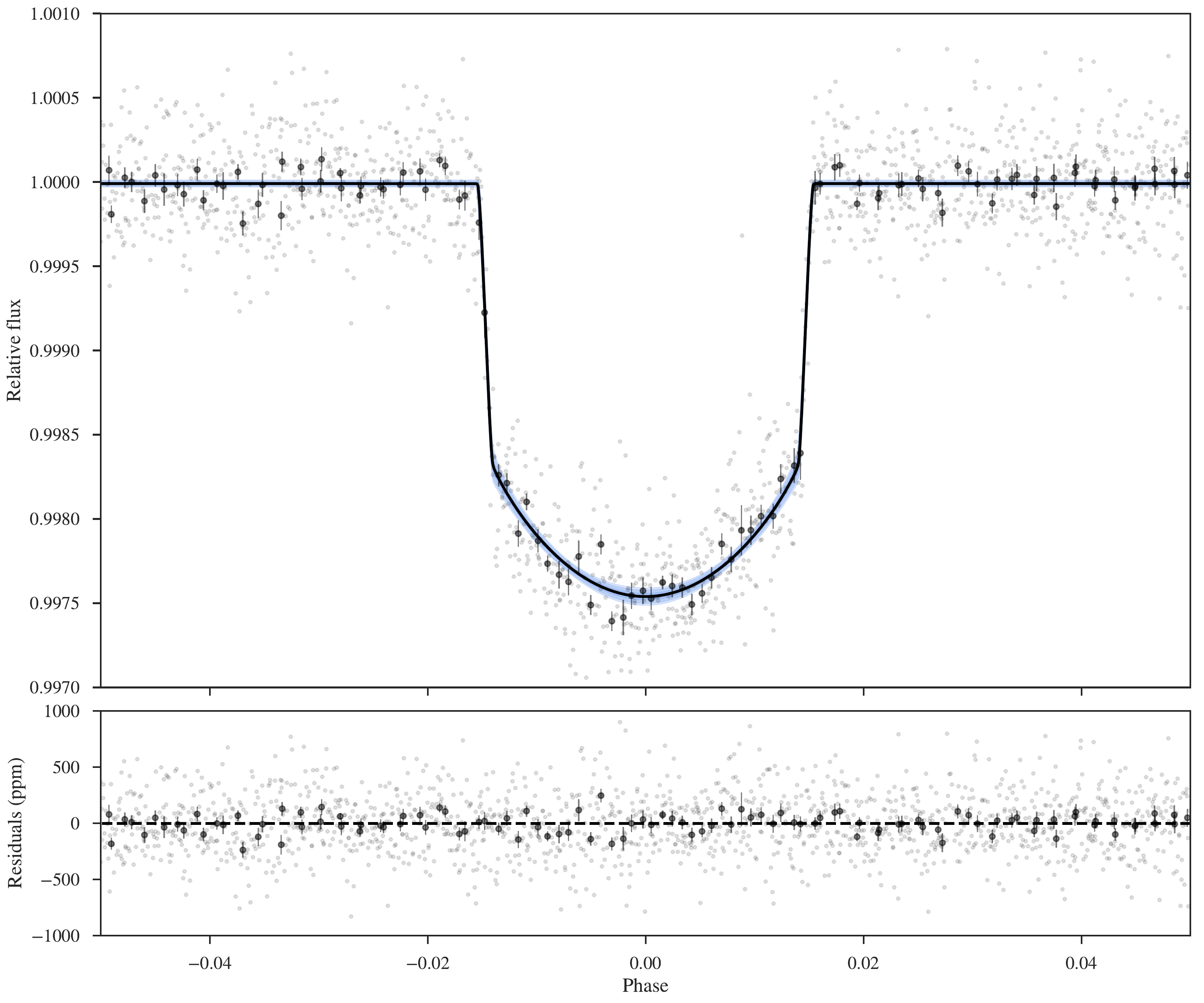}
\caption{Phase folded \tess\ photometry for \stname. The model generated with the derived parameters of our joint modelling is plotted with a black line. The bottom panel shows the corresponding residuals. \label{fig:pht}}
\end{figure}

\begin{figure*}
\plotone{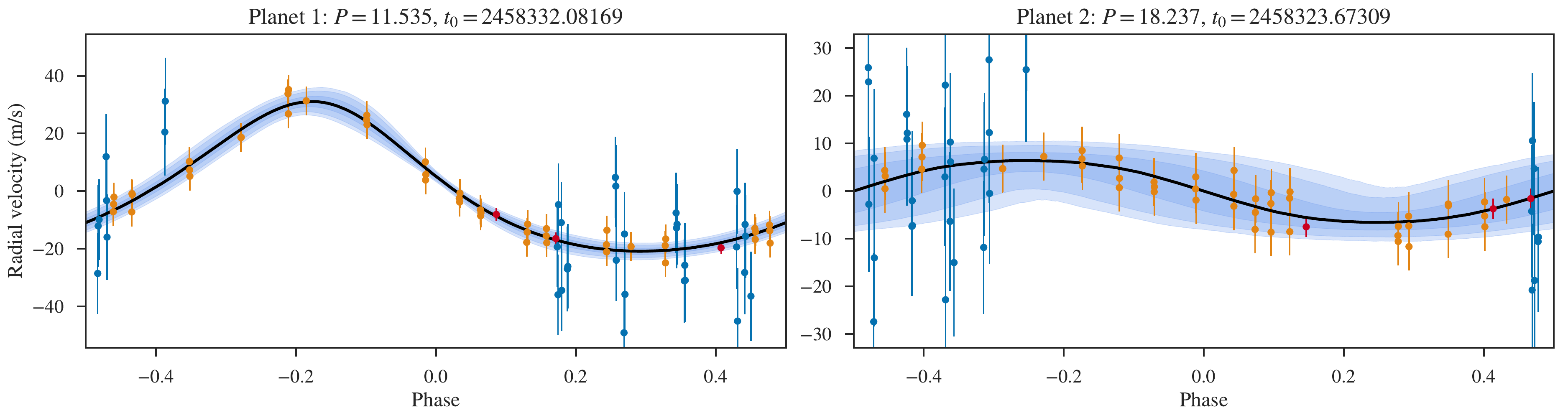}
\caption{Left panel: Radial Velocities as a function of the orbital phase for \plname\ obtained with FIDEOS (blue), FEROS (yellow) and HARPS (red). The black line represents the keplerian model generated from the posterior distributions obtained in Section \ref{sec:anal}. Right panel: same figure but for the second keplerian signal of our modeling.
 \label{fig:phr}}
\end{figure*}

\begin{deluxetable*}{lrc}[b!]
\tablecaption{Planetary properties of the \stname\ system. For the priors, $N(\mu,\sigma)$ stands for a normal distribution with mean $\mu$ and standard deviation $\sigma$, $U(a,b)$ stands for a uniform distribution between $a$ and $b$, and $J(a,b)$ stands for a Jeffrey's prior defined between $a$ and $b$.\label{tab:plprops}}
\tablecolumns{3}
\tablenum{2}
\tablewidth{0pt}
\tablehead{
\colhead{Parameter} &
\colhead{Prior} &
\colhead{Value} \\
}
\startdata
P (days) & $N(11.535,0.001)$  &          \per \\
T$_0$ (BJD)&  $N(2458332.0816,0.001)$&  \TC \\
$a$/R$_{\star}$ & $U(1,300)$ & \ars \\
b & $U(0,1)$ & \imp \\
\rpl/\rstar  & $U(0.01,0.1)$ & \rprs \\
$\sigma_w^{\rm TESS}$ (ppm) & $J(1,500)$ & \sigmatess \\
q$_1^{\rm TESS}$ & $U(0,1)$ & \ldca \\
q$_2^{\rm TESS}$ & $U(0,1)$& \ldcb \\
K (m s$^{-1}$) & $U(0,100)$& \Krv \\
$\sqrt{e}\sin{\omega}$ & $U(-1,1)$ & \sesinw \\
$\sqrt{e}\cos{\omega}$ & $U(-1,1)$ & \secosw \\
$\dot{\gamma}$\ (m s$^{-1}$ / d) & $U(-10,10)$& \rvslope \\
$\gamma_{\rm FEROS}$  (m s$^{-1}$)& $N(30755,10)$& \muferos \\
$\gamma_{\rm HARPS}$ (m s$^{-1}$)& $N(30805,10)$ & \muharps \\
$\gamma_{\rm FIDEOS}$ (m s$^{-1}$)& $N(30699,10)$ & \mufideos \\
P$^c$ (days) & $N(11.54,0.01)$  &  \perc \\
K$^c$ (m s$^{-1}$) & $N(0,100)$& \Krvc \\
T$^c_0$ (BJD)&  $N(2458323.5,3)$&  \TCc \\
\hline
$e$ &  & \ecc \\
$\omega$ (rad) &  & \peri \\
$i$ (deg) &  & 89.1 $\pm$ 0.8 \\
\mpl\ (\mjup)& & \mptess \\
\rpl\ (\rjup)& & \rptess \\
$a$ (AU)    & & \sma \\
\teq\tablenotemark{a}  (K)   & & \teqv \\ 
\enddata
\tablenotetext{a}{Time-averaged equilibrium temperature computed according to equation~16 of \citet{mendez:2017}}
\end{deluxetable*}

\section{Discussion} \label{sec:disc}

\subsection{Structure}
\begin{figure*}
\plotone{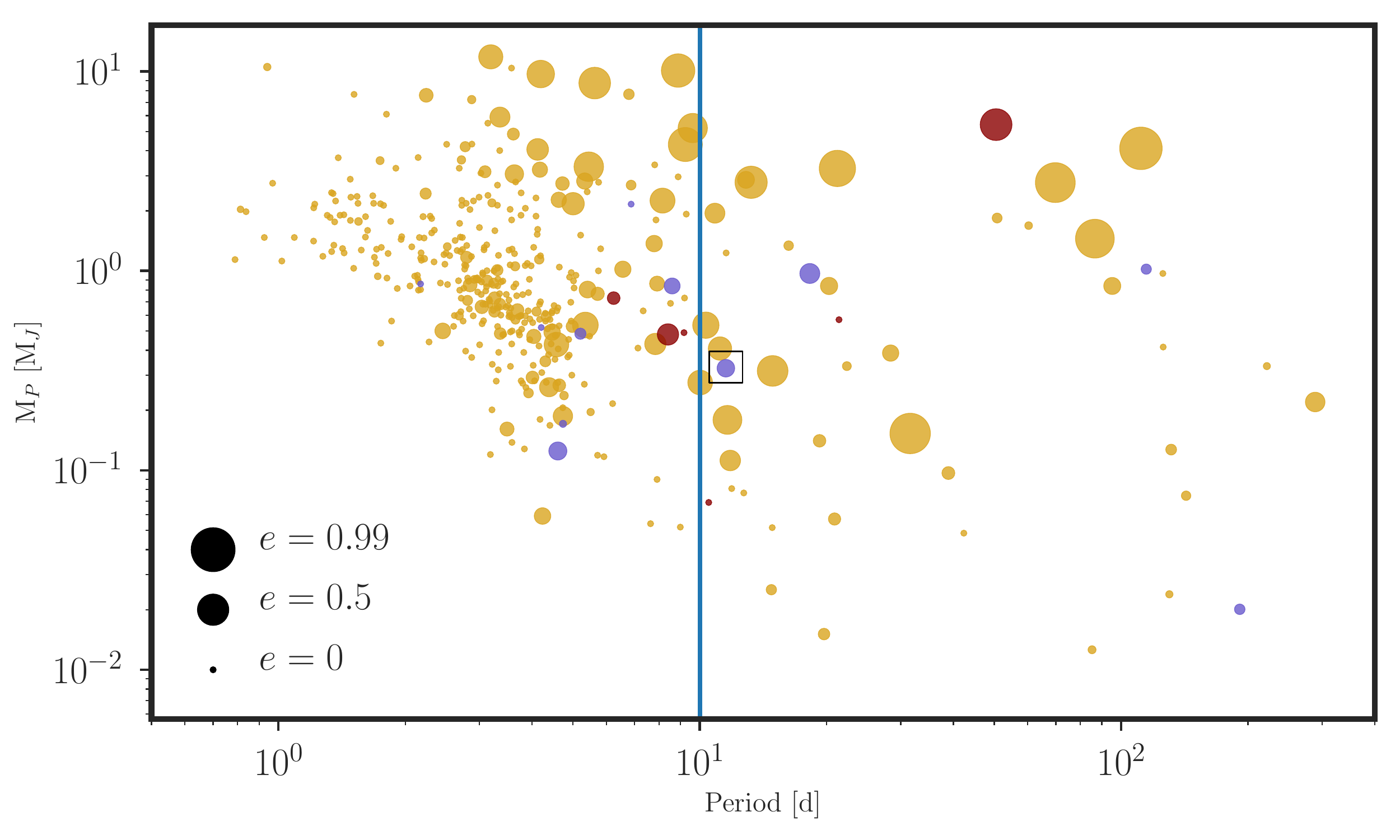}
\caption{Planet mass versus orbital period scatter plot for the population of giant transiting planets with masses and radii measured with a precision of 20\% or better. The size of the points scales with the orbital eccentricity and the color represents the evolutionary state of the host star (yellow: main sequence stars, purple: sub-giant stars, red: giant stars. The vertical line corresponds to the approximate
division between hot and warm Jupiters (10 days). The position of \plname\ is marked with a square.)\label{fig:mp}}
\end{figure*}

\begin{figure*}
\plotone{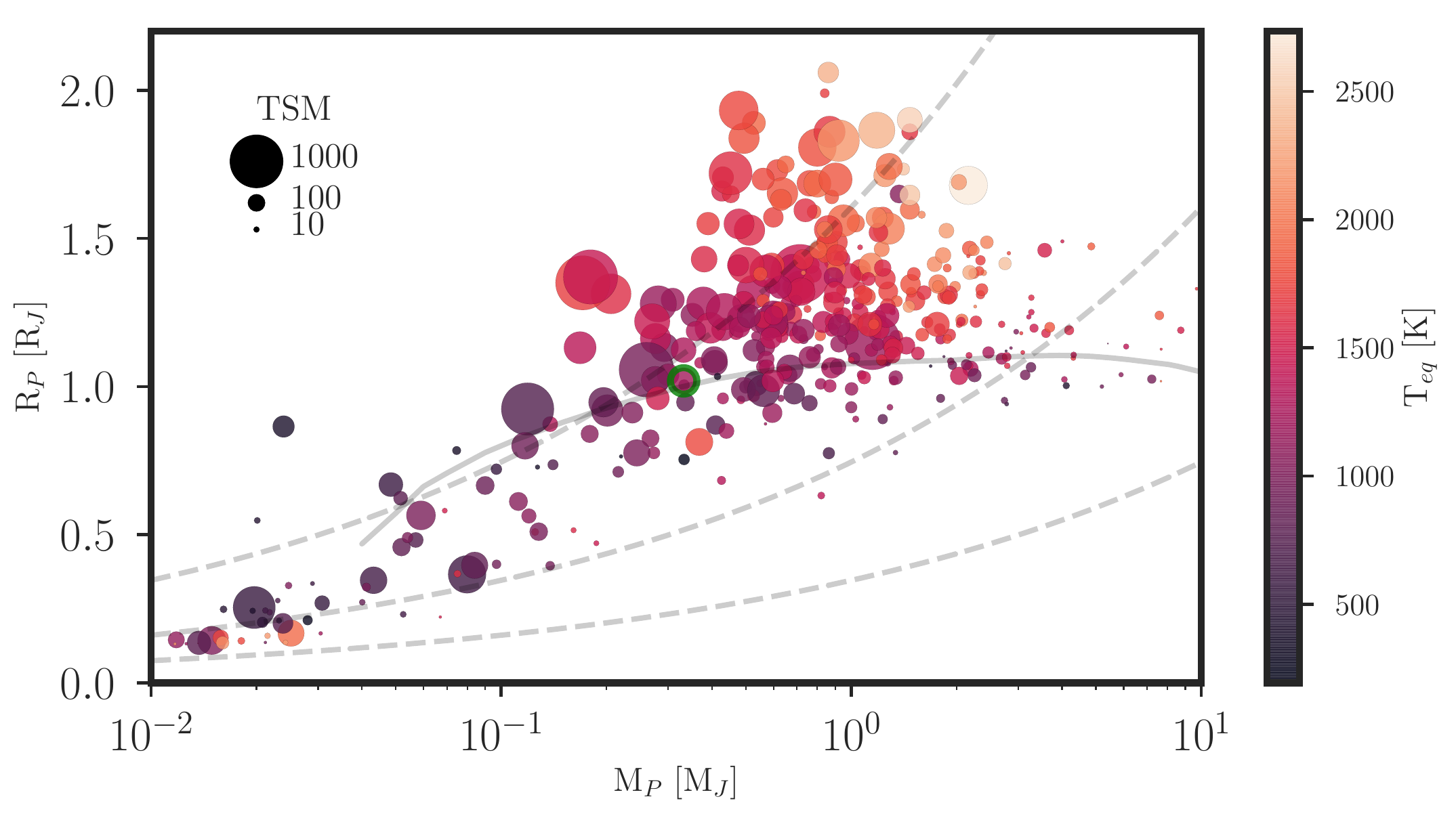}
\caption{Mass -- Radius diagram for the population of well characterized transiting planets. The point corresponding to \plname\ is marked with a green contour. The color represents the equilibrium temperature of the planet, while the size scales down with the transmission spectroscopy metric as defined by \citet{tsm}. The dashed gray lines correspond to iso-density curves for 0.3, 3 and 30 g cm$^{-3}$, respectively. The solid line corresponds to the predicted radius using the models of \citet{fortney:2007} for a planet with a 10~\mearth\ central core.\label{fig:mr}}
\end{figure*}

\plname\ adds up to the still sparse, but recently growing, 
population of transiting giant planets with well constrained parameters
orbiting beyond 0.1 AU (see Figure \ref{fig:mp}).
While located relatively far from its host star if compared to
the typical population of hot Jupiters, the moderately massive stellar
host (\mstar\ = \mst\ \msun) is responsible for producing insolation
levels on \plname\ that are high enough (\teq\ = \teqv\ K) to possibly
alter the internal planetary structure \citep{kovacs:2010, demory:2011}.
Therefore, in terms of structure, \plname\ shares similar properties to
other known low mass hot Jupiters, like WASP-83b \citep{hellier:2014}, HATS-21b
\citep{bhatti:2016}, and WASP-160b \citep{lendl:2018} (see Figure~\ref{fig:mr}).

Regardless of its moderately high equilibrium temperature, the observed
physical properties of \plname\ are still consistent with standard structural
models. Specifically, using the \citet{fortney:2007} models, we obtain that
given the current stellar age, luminosity, and star-planet separation, the
structure of \plname\ requires the presence of a core of
$\approx$8.3 $\pm$ 2.5~\mearth\ of solid material, which is consistent with the values expected in the
core accretion model of planet formation \citep{pollack:96}.

Even though, \plname\ presents a relatively shallow transit depth,
the bright host star coupled to the relatively low density of the planet,
make of the \stname\ system a well suited target for transmission
spectroscopy observations (see Figure~\ref{fig:mr}). Specifically,
with a transmission spectroscopy metric of TSM = 253, \plname\ would
be ranked in the first quartile for atmospheric characterization
as proposed by \citet{tsm}. \stname\ along with HD~209458 \citep{henry:2000}
and HD~189733 \citep{bouchy:2005} are the three brightest systems known to
host transiting giant planets.

\subsection{Orbital Evolution}
The orbital parameters of \plname\ are well
suited for studying the migration mechanism of close-in giant planets.
Its current distance to the star during periastron passages is too
large for producing significant migration by tidal interactions,
and therefore the orbital properties should contain information
about the migration history. As an important fraction of other
recently discovered transiting warm Jupiters, \plname\ presents
a moderate eccentricity ($e$ = \ecc). Eccentricity excitation
is expected to be suppressed during the disk lifetime
\citep{dunhill:2013}, and therefore the gentle disc migration
mechanism is at odds with these observations. Additionally,
planet-planet scattering at these short orbital distances
should preferentially produce collisions between planets,
rather than the excitation of eccentricities\footnote{More quantitatively, the collisions largely dominate as the escape velocity of the planet of $\sim$35km s$^{-1}$ is significantly lower than its orbital velocity of $\sim$100km s$^{-1}$.} \citep{petrovich:2014}.
Eccentricities of the observed population of warm Jupiters
can be enhanced by secular gravitational interactions with
other exterior orbiting companions having eccentric and/or
highly miss-aligned orbits \citep{kozai:62,lidov:62,naoz:2016}.
In this scenario the
eccentricity and spin-orbit angle of the inner planet
suffer periodic variations in timescales significantly
larger than the orbital period, and when the eccentricity
approaches unity, the planets experiences significant
inward migration. Additionally, \citet{dong:2014} predicts
that the perturber should be close enough in order to
overcome the precesion caused by general relativity.
\plname\ presents properties that can be associated with this last
scenario of planet migration. In addition to its eccentric orbit,
\plname\ experiences an acceleration that could be associated
with the presence of a massive outer companion, a scenario that can be
tested with long term radial velocity monitoring. Additionally,
the measurement of the miss-alignment angle between the spin of the
star and the orbital plane through the observation of the Rossiter-
McLaughlin effect, could further help in constraining migration mechanism.
While the expected amplitude of the Rossiter-McLaughlin effect is relatively small
(K$_{RM} \approx 8$ m s$^{-1}$ for an aligned orbit), the signal
should be detectable by a wide range of instruments given the
brightness of the host star.

It is interesting also to note that the second signal observed
in the radial velocities can be probably ruled out as having a
planetary origin based on stability criteria. For example,
according to the semi-empirical criterion by \citet{giuppone:2013},
a planet with a mass of 0.12 \mjup on a circular orbit with a period
of 18.24 d, would
make the system unstable as
\begin{equation}
    1 - \left(\frac{P}{P^{c}}\right)^{2/3} \frac{1+e}{1-e^{c}} < 1.57 \bigg[ \left(\frac{M_P}{\mstar}\right)^{2/7} + \left(\frac{M_P^c}{\mstar}\right)^{2/7} \bigg].
\end{equation}
We caution, however, that the candidate has a period ratio of
$\approx$1.58, which places the planet pair near to the 5:3
mean-motion resonance, possibly allowing for islands of stability.
Assessing the long-term stability of the system is beyond the scope
of this work.

\subsection{Evolved host star}
\plname\ is a peculiar system due to the evolutionary stage
of its host star. Radial velocity surveys have found that
close-in giant planets are rare around giant and sub-giant
stars, if compared to the the occurrence rate on main sequence
stars \citep[e.g.][]{bowler:2010,johnson:2010,Jones:2014}.
Two mechanisms have been invoked to explain this.
First, the inferred stellar masses for this
sample are systematically higher than those of the giant planets
orbiting main sequence stars, and hence the formation of giant planets at close orbital distances or their migration from beyond the snowline could be prevented due to
a shorter lifetime of the protoplanetary disc \citep[e.g.][]{currie:2009}.
On the other hand, the scarcity of giant planets orbiting evolved stars
could be also produced by
the spiral decay of the orbits due to the increase in the
strength of stellar tides \citep[e.g.][]{villaver:2009, schlaufman:2013, villaver:2014}. 
\plname\ joins the small group of $\approx$~10 well characterized
transiting systems of giant planets in close-in orbits around sub-giant
stars (see Figure \ref{fig:mp}). An important fraction of these close-in systems,
including \plname, present accelerations in their radial velocity curves:
Kepler-435b \citep{almenara:2015},
K2-39b \citep{vaneylen:2016},
K2-99b \citep{smith:2017},
KELT-11b \citep{pepper:2017}.
While \citet{knutson:2014} shows that 50\% of the population
of hot Jupiters orbiting main sequence stars have outer companions,
the radial velocity slopes derived for this population are
statistically smaller than those present in the population
of close-in planets orbiting sub-giant stars. These observations
indicate that the populations of close-in planets orbiting main sequence
and sub-giant stars are probably distinct, where the latter have more
massive and/or closer companions.
In the forthcoming years, the
\tess\ mission is expected to produce a statistically significant
sample of close-in transiting giant planets orbiting evolved stars,
which will allow to further study the origin of the low occurrence rate of this
type of planetary systems.

\acknowledgements
R.B.\ acknowledges support from FONDECYT Post-doctoral Fellowship Project 3180246, and from the Millennium Institute of Astrophysics (MAS).
A.J.\ acknowledges support from FONDECYT project 1171208, CONICYT project Basal AFB-170002, and by the Ministry for the Economy, Development, and Tourism's Programa Iniciativa Cient\'{i}fica Milenio through grant IC\,120009, awarded to the Millennium Institute of Astrophysics (MAS). 
M.R.D. is supported by CONICYT-PFCHA/Doctorado Nacional-21140646, Chile.
J.S.J. acknowledges support by FONDECYT project 1161218 and partial
support by BASAL CATA PFB-06.
A.Z.\ acknowledges support by CONICYT-PFCHA/Doctorado
Nacional 21170536, Chile. 
%
We acknowledge the use of \tess\ Alert data, which is currently in a beta test phase, from pipelines at the \tess\ Science Office and at the \tess\ Science Processing Operations Center.
This research has made use of the Exoplanet Follow-up Observation Program website, which is operated by the California Institute of Technology, under contract with the National Aeronautics and Space Administration under the Exoplanet Exploration Program.
This paper includes data collected by the \tess\ mission, which are publicly available from the Mikulski Archive for Space Telescopes (MAST).
We thank Sam Kim, R\'egis Lachaume and Martin Schlecker for
their technical assistance during the observations
at the MPG 2.2 m Telescope




\bibliography{k2clbib}

\begin{thebibliography}{}
\expandafter\ifx\csname natexlab\endcsname\relax\def\natexlab#1{#1}\fi
\providecommand{\url}[1]{\href{#1}{#1}}

\bibitem[{{Aguilera-G{\'o}mez} {et~al.}(2016){Aguilera-G{\'o}mez},
  {Chanam{\'e}}, {Pinsonneault}, \& {Carlberg}}]{aguilera:2016}
{Aguilera-G{\'o}mez}, C., {Chanam{\'e}}, J., {Pinsonneault}, M.~H., \&
  {Carlberg}, J.~K. 2016, \apj, 829, 127

\bibitem[{{Almenara} {et~al.}(2015){Almenara}, {Damiani}, {Bouchy}, {Havel},
  {Bruno}, {H{\'e}brard}, {Diaz}, {Deleuil}, {Barros}, {Boisse}, {Bonomo},
  {Montagnier}, \& {Santerne}}]{almenara:2015}
{Almenara}, J.~M., {Damiani}, C., {Bouchy}, F., {et~al.} 2015, \aap, 575, A71

\bibitem[{{Anderson} \& {Lai}(2017)}]{anderson:2017}
{Anderson}, K.~R., \& {Lai}, D. 2017, \mnras, 472, 3692

\bibitem[{{Bakos} {et~al.}(2004){Bakos}, {Noyes}, {Kov{\'a}cs}, {Stanek},
  {Sasselov}, \& {Domsa}}]{bakos:2004}
{Bakos}, G., {Noyes}, R.~W., {Kov{\'a}cs}, G., {et~al.} 2004, \pasp, 116, 266

\bibitem[{{Bakos} {et~al.}(2013){Bakos}, {Csubry}, {Penev}, {Bayliss},
  {Jord{\'a}n}, {Afonso}, {Hartman}, {Henning}, {Kov{\'a}cs}, {Noyes},
  {B{\'e}ky}, {Suc}, {Cs{\'a}k}, {Rabus}, {L{\'a}z{\'a}r}, {Papp}, {S{\'a}ri},
  {Conroy}, {Zhou}, {Sackett}, {Schmidt}, {Mancini}, {Sasselov}, \&
  {Ueltzhoeffer}}]{bakos:2013}
{Bakos}, G.~{\'A}., {Csubry}, Z., {Penev}, K., {et~al.} 2013, \pasp, 125, 154

\bibitem[{{Baraffe} {et~al.}(2015){Baraffe}, {Homeier}, {Allard}, \&
  {Chabrier}}]{baraffe:2015}
{Baraffe}, I., {Homeier}, D., {Allard}, F., \& {Chabrier}, G. 2015, \aap, 577,
  A42

\bibitem[{{Barclay} {et~al.}(2018){Barclay}, {Pepper}, \&
  {Quintana}}]{barclay:2018}
{Barclay}, T., {Pepper}, J., \& {Quintana}, E.~V. 2018, ArXiv e-prints,
  arXiv:1804.05050

\bibitem[{{Bento} {et~al.}(2018){Bento}, {Hartman}, {Bakos}, {Bhatti},
  {Csubry}, {Penev}, {Bayliss}, {de Val-Borro}, {Zhou}, {Brahm}, {Espinoza},
  {Rabus}, {Jord{\'a}n}, {Suc}, {Ciceri}, {Sarkis}, {Henning}, {Mancini},
  {Tinney}, {Wright}, {Durkan}, {Tan}, {L{\'a}z{\'a}r}, {Papp}, \&
  {S{\'a}ri}}]{bento:2018}
{Bento}, J., {Hartman}, J.~D., {Bakos}, G.~{\'A}., {et~al.} 2018, \mnras, 477,
  3406

\bibitem[{{Bhatti} {et~al.}(2016){Bhatti}, {Bakos}, {Hartman}, {Zhou}, {Penev},
  {Bayliss}, {Jord{\'a}n}, {Brahm}, {Espinoza}, {Rabus}, {Mancini}, {de
  Val-Borro}, {Bento}, {Ciceri}, {Csubry}, {Henning}, {Schmidt}, {Arriagada},
  {Butler}, {Crane}, {Shectman}, {Thompson}, {Tan}, {Suc}, {L{\'a}z{\'a}r},
  {Papp}, \& {S{\'a}ri}}]{bhatti:2016}
{Bhatti}, W., {Bakos}, G.~{\'A}., {Hartman}, J.~D., {et~al.} 2016, ArXiv
  e-prints, arXiv:1607.00322

\bibitem[{{Bouchy} {et~al.}(2005){Bouchy}, {Udry}, {Mayor}, {Moutou}, {Pont},
  {Iribarne}, {da Silva}, {Ilovaisky}, {Queloz}, {Santos}, {S{\'e}gransan}, \&
  {Zucker}}]{bouchy:2005}
{Bouchy}, F., {Udry}, S., {Mayor}, M., {et~al.} 2005, \aap, 444, L15

\bibitem[{{Bowler} {et~al.}(2010){Bowler}, {Johnson}, {Marcy}, {Henry}, {Peek},
  {Fischer}, {Clubb}, {Liu}, {Reffert}, {Schwab}, \& {Lowe}}]{bowler:2010}
{Bowler}, B.~P., {Johnson}, J.~A., {Marcy}, G.~W., {et~al.} 2010, \apj, 709,
  396

\bibitem[{{Brahm} {et~al.}(2017{\natexlab{a}}){Brahm}, {Jord{\'a}n}, \&
  {Espinoza}}]{brahm:2017:ceres}
{Brahm}, R., {Jord{\'a}n}, A., \& {Espinoza}, N. 2017{\natexlab{a}}, \pasp,
  129, 034002

\bibitem[{{Brahm} {et~al.}(2017{\natexlab{b}}){Brahm}, {Jord{\'a}n}, {Hartman},
  \& {Bakos}}]{brahm:2016:zaspe}
{Brahm}, R., {Jord{\'a}n}, A., {Hartman}, J., \& {Bakos}, G.
  2017{\natexlab{b}}, \mnras, 467, 971

\bibitem[{{Brahm} {et~al.}(2016){Brahm}, {Jord{\'a}n}, {Bakos}, {Penev},
  {Espinoza}, {Rabus}, {Hartman}, {Bayliss}, {Ciceri}, {Zhou}, {Mancini},
  {Tan}, {de Val-Borro}, {Bhatti}, {Csubry}, {Bento}, {Henning}, {Schmidt},
  {Rojas}, {Suc}, {L{\'a}z{\'a}r}, {Papp}, \& {S{\'a}ri}}]{brahm:2016:hs17}
{Brahm}, R., {Jord{\'a}n}, A., {Bakos}, G.~{\'A}., {et~al.} 2016, \aj, 151, 89

\bibitem[{{Brahm} {et~al.}(2018){Brahm}, {Espinoza}, {Jord{\'a}n}, {Rojas},
  {Sarkis}, {D{\'{\i}}az}, {Rabus}, {Drass}, {Lachaume}, {Soto}, {Jenkins},
  {Jones}, {Henning}, {Pantoja}, \& {Vu{\v c}kovi{\'c}}}]{k2-232}
{Brahm}, R., {Espinoza}, N., {Jord{\'a}n}, A., {et~al.} 2018, \mnras, 477, 2572

\bibitem[{{Buchner} {et~al.}(2014){Buchner}, {Georgakakis}, {Nandra}, {Hsu},
  {Rangel}, {Brightman}, {Merloni}, {Salvato}, {Donley}, \&
  {Kocevski}}]{PyMultiNest}
{Buchner}, J., {Georgakakis}, A., {Nandra}, K., {et~al.} 2014, \aap, 564, A125

\bibitem[{{Castelli} \& {Kurucz}(2004)}]{atlas9}
{Castelli}, F., \& {Kurucz}, R.~L. 2004, ArXiv e-prints, astro

\bibitem[{{Chen} {et~al.}(2018){Chen}, {Pall{\'e}}, {Welbanks},
  {Prieto-Arranz}, {Madhusudhan}, {Gandhi}, {Casasayas-Barris}, {Murgas},
  {Nortmann}, {Crouzet}, {Parviainen}, \& {Gandolfi}}]{chen:2018}
{Chen}, G., {Pall{\'e}}, E., {Welbanks}, L., {et~al.} 2018, \aap, 616, A145

\bibitem[{{Currie}(2009)}]{currie:2009}
{Currie}, T. 2009, \apjl, 694, L171

\bibitem[{{Demory} \& {Seager}(2011)}]{demory:2011}
{Demory}, B.-O., \& {Seager}, S. 2011, \apjs, 197, 12

\bibitem[{{Dong} {et~al.}(2014){Dong}, {Katz}, \& {Socrates}}]{dong:2014}
{Dong}, S., {Katz}, B., \& {Socrates}, A. 2014, \apjl, 781, L5

\bibitem[{{Dunhill} {et~al.}(2013){Dunhill}, {Alexander}, \&
  {Armitage}}]{dunhill:2013}
{Dunhill}, A.~C., {Alexander}, R.~D., \& {Armitage}, P.~J. 2013, \mnras, 428,
  3072

\bibitem[{{Espinoza} {et~al.}(2016){Espinoza}, {Brahm}, {Jord{\'a}n},
  {Jenkins}, {Rojas}, {Jofr{\'e}}, {M{\"a}dler}, {Rabus}, {Chanam{\'e}},
  {Pantoja}, {Soto}, {Morzinski}, {Males}, {Ward-Duong}, \&
  {Close}}]{espinoza:2016:exo}
{Espinoza}, N., {Brahm}, R., {Jord{\'a}n}, A., {et~al.} 2016, \apj, 830, 43

\bibitem[{{Esposito} {et~al.}(2017){Esposito}, {Covino}, {Desidera}, {Mancini},
  {Nascimbeni}, {Zanmar Sanchez}, {Biazzo}, {Lanza}, {Leto}, {Southworth},
  {Bonomo}, {Su{\'a}rez Mascare{\~n}o}, {Boccato}, {Cosentino}, {Claudi},
  {Gratton}, {Maggio}, {Micela}, {Molinari}, {Pagano}, {Piotto}, {Poretti},
  {Smareglia}, {Sozzetti}, {Affer}, {Anderson}, {Andreuzzi}, {Benatti},
  {Bignamini}, {Borsa}, {Borsato}, {Ciceri}, {Damasso}, {di Fabrizio},
  {Giacobbe}, {Granata}, {Harutyunyan}, {Henning}, {Malavolta}, {Maldonado},
  {Martinez Fiorenzano}, {Masiero}, {Molaro}, {Molinaro}, {Pedani}, {Rainer},
  {Scandariato}, \& {Turner}}]{esposito:2017}
{Esposito}, M., {Covino}, E., {Desidera}, S., {et~al.} 2017, \aap, 601, A53

\bibitem[{{Feroz} {et~al.}(2009){Feroz}, {Hobson}, \& {Bridges}}]{MultiNest}
{Feroz}, F., {Hobson}, M.~P., \& {Bridges}, M. 2009, \mnras, 398, 1601

\bibitem[{{Foreman-Mackey} {et~al.}(2017){Foreman-Mackey}, {Agol},
  {Ambikasaran}, \& {Angus}}]{celerite}
{Foreman-Mackey}, D., {Agol}, E., {Ambikasaran}, S., \& {Angus}, R. 2017, \aj,
  154, 220

\bibitem[{{Foreman-Mackey} {et~al.}(2013){Foreman-Mackey}, {Hogg}, {Lang}, \&
  {Goodman}}]{emcee:2013}
{Foreman-Mackey}, D., {Hogg}, D.~W., {Lang}, D., \& {Goodman}, J. 2013, \pasp,
  125, 306

\bibitem[{{Fortney} {et~al.}(2007){Fortney}, {Marley}, \&
  {Barnes}}]{fortney:2007}
{Fortney}, J.~J., {Marley}, M.~S., \& {Barnes}, J.~W. 2007, \apj, 659, 1661

\bibitem[{{Fulton} {et~al.}(2018){Fulton}, {Petigura}, {Blunt}, \&
  {Sinukoff}}]{fulton:2018}
{Fulton}, B.~J., {Petigura}, E.~A., {Blunt}, S., \& {Sinukoff}, E. 2018, ArXiv
  e-prints, arXiv:1801.01947

\bibitem[{{Gaia Collaboration} {et~al.}(2018){Gaia Collaboration}, {Brown},
  {Vallenari}, {Prusti}, {de Bruijne}, {Babusiaux}, \&
  {Bailer-Jones}}]{gaia:dr2}
{Gaia Collaboration}, {Brown}, A.~G.~A., {Vallenari}, A., {et~al.} 2018, ArXiv
  e-prints, arXiv:1804.09365

\bibitem[{{Gaia Collaboration} {et~al.}(2016){Gaia Collaboration}, {Prusti},
  {de Bruijne}, {Brown}, {Vallenari}, {Babusiaux}, {Bailer-Jones}, {Bastian},
  {Biermann}, {Evans}, \& et~al.}]{gaia}
{Gaia Collaboration}, {Prusti}, T., {de Bruijne}, J.~H.~J., {et~al.} 2016,
  \aap, 595, A1

\bibitem[{{Giuppone} {et~al.}(2013){Giuppone}, {Morais}, \&
  {Correia}}]{giuppone:2013}
{Giuppone}, C.~A., {Morais}, M.~H.~M., \& {Correia}, A.~C.~M. 2013, \mnras,
  436, 3547

\bibitem[{{Grunblatt} {et~al.}(2017){Grunblatt}, {Huber}, {Gaidos}, {Lopez},
  {Howard}, {Isaacson}, {Sinukoff}, {Vanderburg}, {Nofi}, {Yu}, {North},
  {Chaplin}, {Foreman-Mackey}, {Petigura}, {Ansdell}, {Weiss}, {Fulton}, \&
  {Lin}}]{grunblatt:2017}
{Grunblatt}, S.~K., {Huber}, D., {Gaidos}, E., {et~al.} 2017, \aj, 154, 254

\bibitem[{{Hartman} {et~al.}(2012){Hartman}, {Bakos}, {B{\'e}ky}, {Torres},
  {Latham}, {Csubry}, {Penev}, {Shporer}, {Fulton}, {Buchhave}, {Johnson},
  {Howard}, {Marcy}, {Fischer}, {Kov{\'a}cs}, {Noyes}, {Esquerdo}, {Everett},
  {Szklen{\'a}r}, {Quinn}, {Bieryla}, {Knox}, {Hinz}, {Sasselov}, {F{\H
  u}r{\'e}sz}, {Stefanik}, {L{\'a}z{\'a}r}, {Papp}, \&
  {S{\'a}ri}}]{hartman:2012}
{Hartman}, J.~D., {Bakos}, G.~{\'A}., {B{\'e}ky}, B., {et~al.} 2012, \aj, 144,
  139

\bibitem[{{Hellier} {et~al.}(2015){Hellier}, {Anderson}, {Collier Cameron},
  {Delrez}, {Gillon}, {Jehin}, {Lendl}, {Maxted}, {Pepe}, {Pollacco}, {Queloz},
  {S{\'e}gransan}, {Smalley}, {Smith}, {Southworth}, {Triaud}, {Turner},
  {Udry}, \& {West}}]{hellier:2014}
{Hellier}, C., {Anderson}, D.~R., {Collier Cameron}, A., {et~al.} 2015, \aj,
  150, 18

\bibitem[{{Hellier} {et~al.}(2017){Hellier}, {Anderson}, {Cameron}, {Delrez},
  {Gillon}, {Jehin}, {Lendl}, {Maxted}, {Neveu-VanMalle}, {Pepe}, {Pollacco},
  {Queloz}, {S{\'e}gransan}, {Smalley}, {Southworth}, {Triaud}, {Udry}, {Wagg},
  \& {West}}]{wasp130}
{Hellier}, C., {Anderson}, D.~R., {Cameron}, A.~C., {et~al.} 2017, \mnras, 465,
  3693

\bibitem[{{Henry} {et~al.}(2000){Henry}, {Marcy}, {Butler}, \&
  {Vogt}}]{henry:2000}
{Henry}, G.~W., {Marcy}, G.~W., {Butler}, R.~P., \& {Vogt}, S.~S. 2000, \apjl,
  529, L41

\bibitem[{{Howell} {et~al.}(2014){Howell}, {Sobeck}, {Haas}, {Still},
  {Barclay}, {Mullally}, {Troeltzsch}, {Aigrain}, {Bryson}, {Caldwell},
  {Chaplin}, {Cochran}, {Huber}, {Marcy}, {Miglio}, {Najita}, {Smith},
  {Twicken}, \& {Fortney}}]{howell:2014}
{Howell}, S.~B., {Sobeck}, C., {Haas}, M., {et~al.} 2014, Publications of the
  Astronomical Society of the Pacific, 126, 398

\bibitem[{{Hussain}(2002)}]{hussain:2002}
{Hussain}, G.~A.~J. 2002, Astronomische Nachrichten, 323, 349

\bibitem[{{Jenkins} {et~al.}(2008){Jenkins}, {Jones}, {Pavlenko}, {Pinfield},
  {Barnes}, \& {Lyubchik}}]{jenkins:2008}
{Jenkins}, J.~S., {Jones}, H.~R.~A., {Pavlenko}, Y., {et~al.} 2008, \aap, 485,
  571

\bibitem[{{Jensen} {et~al.}(2018){Jensen}, {Cauley}, {Redfield}, {Cochran}, \&
  {Endl}}]{jensen:2018}
{Jensen}, A.~G., {Cauley}, P.~W., {Redfield}, S., {Cochran}, W.~D., \& {Endl},
  M. 2018, \aj, 156, 154

\bibitem[{{Johnson} {et~al.}(2010){Johnson}, {Howard}, {Bowler}, {Henry},
  {Marcy}, {Wright}, {Fischer}, \& {Isaacson}}]{johnson:2010}
{Johnson}, J.~A., {Howard}, A.~W., {Bowler}, B.~P., {et~al.} 2010, \pasp, 122,
  701

\bibitem[{{Jones} {et~al.}(2017){Jones}, {Brahm}, {Wittenmyer}, {Drass},
  {Jenkins}, {Melo}, {Vos}, \& {Rojo}}]{jones:2017}
{Jones}, M.~I., {Brahm}, R., {Wittenmyer}, R.~A., {et~al.} 2017, \aap, 602, A58

\bibitem[{{Jones} {et~al.}(2014){Jones}, {Jenkins}, {Bluhm}, {Rojo}, \&
  {Melo}}]{Jones:2014}
{Jones}, M.~I., {Jenkins}, J.~S., {Bluhm}, P., {Rojo}, P., \& {Melo}, C.~H.~F.
  2014, \aap, 566, A113

\bibitem[{{Jones} {et~al.}(2018){Jones}, {Brahm}, {Espinoza}, {Jord{\'a}n},
  {Rojas}, {Rabus}, {Drass}, {Zapata}, {Soto}, {Jenkins}, {Vu{\v c}kovi{\'c}},
  {Ciceri}, \& {Sarkis}}]{jones:2018}
{Jones}, M.~I., {Brahm}, R., {Espinoza}, N., {et~al.} 2018, \aap, 613, A76

\bibitem[{{Jord{\'a}n} {et~al.}(2018){Jord{\'a}n}, {Brahm}, {Espinoza},
  {Cort{\'e}s}, {D{\'{\i}}az}, {Drass}, {Henning}, {Jenkins}, {Jones}, {Rabus},
  {Rojas}, {Sarkis}, {Vu{\v c}kovi{\'c}}, {Zapata}, {Soto}, {Bakos}, {Bayliss},
  {Bhatti}, {Csubry}, {Lachaume}, {Moraga}, {Pantoja}, {Osip}, {Shporer},
  {Suc}, \& {V{\'a}squez}}]{jordan:2018}
{Jord{\'a}n}, A., {Brahm}, R., {Espinoza}, N., {et~al.} 2018, ArXiv e-prints,
  arXiv:1809.08879

\bibitem[{{Kaufer} {et~al.}(1999){Kaufer}, {Stahl}, {Tubbesing},
  {N{\o}rregaard}, {Avila}, {Francois}, {Pasquini}, \& {Pizzella}}]{kaufer:99}
{Kaufer}, A., {Stahl}, O., {Tubbesing}, S., {et~al.} 1999, The Messenger, 95, 8

\bibitem[{{Kempton} {et~al.}(2018){Kempton}, {Bean}, {Louie}, {Deming}, {Koll},
  {Mansfield}, {Lopez-Morales}, {Swain}, {Zellem}, {Ballard}, {Barclay},
  {Barstow}, {Batalha}, {Beatty}, {Berta-Thompson}, {Birkby}, {Buchhave},
  {Charbonneau}, {Christiansen}, {Cowan}, {Crossfield}, {de Val-Borro},
  {Doyon}, {Dragomir}, {Gaidos}, {Heng}, {Kane}, {Kreidberg}, {Mallonn},
  {Morley}, {Narita}, {Nascimbeni}, {Palle}, {Quintana}, {Rauscher}, {Seager},
  {Shkolnik}, {Sing}, {Sozzetti}, {Stassun}, {Valenti}, \& {von Essen}}]{tsm}
{Kempton}, E.~M.-R., {Bean}, J.~L., {Louie}, D.~R., {et~al.} 2018, ArXiv
  e-prints, arXiv:1805.03671

\bibitem[{{Kipping}(2013)}]{Kipping:LDs}
{Kipping}, D.~M. 2013, \mnras, 435, 2152

\bibitem[{{Knutson} {et~al.}(2014){Knutson}, {Fulton}, {Montet}, {Kao}, {Ngo},
  {Howard}, {Crepp}, {Hinkley}, {Bakos}, {Batygin}, {Johnson}, {Morton}, \&
  {Muirhead}}]{knutson:2014}
{Knutson}, H.~A., {Fulton}, B.~J., {Montet}, B.~T., {et~al.} 2014, \apj, 785,
  126

\bibitem[{{Kov{\'a}cs} {et~al.}(2010){Kov{\'a}cs}, {Bakos}, {Hartman},
  {Torres}, {Noyes}, {Latham}, {Howard}, {Fischer}, {Johnson}, {Marcy},
  {Isaacson}, {Sasselov}, {Stefanik}, {Esquerdo}, {Fernandez}, {L{\'a}z{\'a}r},
  {Papp}, \& {S{\'a}ri}}]{kovacs:2010}
{Kov{\'a}cs}, G., {Bakos}, G.~{\'A}., {Hartman}, J.~D., {et~al.} 2010, \apj,
  724, 866

\bibitem[{{Kozai}(1962)}]{kozai:62}
{Kozai}, Y. 1962, \aj, 67, 591

\bibitem[{{Kreidberg}(2015)}]{kreidberg:2015}
{Kreidberg}, L. 2015, \pasp, 127, 1161

\bibitem[{{Lendl} {et~al.}(2014){Lendl}, {Triaud}, {Anderson}, {Collier
  Cameron}, {Delrez}, {Doyle}, {Gillon}, {Hellier}, {Jehin}, {Maxted},
  {Neveu-VanMalle}, {Pepe}, {Pollacco}, {Queloz}, {S{\'e}gransan}, {Smalley},
  {Smith}, {Udry}, {Van Grootel}, \& {West}}]{wasp117}
{Lendl}, M., {Triaud}, A.~H.~M.~J., {Anderson}, D.~R., {et~al.} 2014, \aap,
  568, A81

\bibitem[{{Lendl} {et~al.}(2018){Lendl}, {Anderson}, {Bonfanti}, {Bouchy},
  {Burdanov}, {Collier Cameron}, {Delrez}, {Gillon}, {Hellier}, {Jehin},
  {Maxted}, {Nielsen}, {Pepe}, {Pollacco}, {Queloz}, {S{\'e}gransan},
  {Southworth}, {Smalley}, {Thompson}, {Turner}, {Triaud}, {Udry}, \&
  {West}}]{lendl:2018}
{Lendl}, M., {Anderson}, D.~R., {Bonfanti}, A., {et~al.} 2018, \mnras,
  arXiv:1807.06973

\bibitem[{{Lidov}(1962)}]{lidov:62}
{Lidov}, M.~L. 1962, \planss, 9, 719

\bibitem[{{Lopez} \& {Fortney}(2016)}]{lopez:2016}
{Lopez}, E.~D., \& {Fortney}, J.~J. 2016, \apj, 818, 4

\bibitem[{{Mayor} {et~al.}(2003){Mayor}, {Pepe}, {Queloz}, {Bouchy},
  {Rupprecht}, {Lo Curto}, {Avila}, {Benz}, {Bertaux}, {Bonfils}, {Dall},
  {Dekker}, {Delabre}, {Eckert}, {Fleury}, {Gilliotte}, {Gojak}, {Guzman},
  {Kohler}, {Lizon}, {Longinotti}, {Lovis}, {Megevand}, {Pasquini}, {Reyes},
  {Sivan}, {Sosnowska}, {Soto}, {Udry}, {van Kesteren}, {Weber}, \&
  {Weilenmann}}]{mayor:2003}
{Mayor}, M., {Pepe}, F., {Queloz}, D., {et~al.} 2003, The Messenger, 114, 20

\bibitem[{{M{\'e}ndez} \& {Rivera-Valent{\'{\i}}n}(2017)}]{mendez:2017}
{M{\'e}ndez}, A., \& {Rivera-Valent{\'{\i}}n}, E.~G. 2017, \apjl, 837, L1

\bibitem[{{Naoz}(2016)}]{naoz:2016}
{Naoz}, S. 2016, \araa, 54, 441

\bibitem[{{Pepper} {et~al.}(2007){Pepper}, {Pogge}, {DePoy}, {Marshall},
  {Stanek}, {Stutz}, {Poindexter}, {Siverd}, {O'Brien}, {Trueblood}, \&
  {Trueblood}}]{pepper:2007}
{Pepper}, J., {Pogge}, R.~W., {DePoy}, D.~L., {et~al.} 2007, \pasp, 119, 923

\bibitem[{{Pepper} {et~al.}(2017){Pepper}, {Rodriguez}, {Collins}, {Johnson},
  {Fulton}, {Howard}, {Beatty}, {Stassun}, {Isaacson}, {Col{\'o}n}, {Lund},
  {Kuhn}, {Siverd}, {Gaudi}, {Tan}, {Curtis}, {Stockdale}, {Mawet}, {Bottom},
  {James}, {Zhou}, {Bayliss}, {Cargile}, {Bieryla}, {Penev}, {Latham},
  {Labadie-Bartz}, {Kielkopf}, {Eastman}, {Oberst}, {Jensen}, {Nelson},
  {Sliski}, {Wittenmyer}, {McCrady}, {Wright}, {Relles}, {Stevens}, {Joner}, \&
  {Hintz}}]{pepper:2017}
{Pepper}, J., {Rodriguez}, J.~E., {Collins}, K.~A., {et~al.} 2017, \aj, 153,
  215

\bibitem[{{Petrovich} \& {Tremaine}(2016)}]{petrovich:2016}
{Petrovich}, C., \& {Tremaine}, S. 2016, \apj, 829, 132

\bibitem[{{Petrovich} {et~al.}(2014){Petrovich}, {Tremaine}, \&
  {Rafikov}}]{petrovich:2014}
{Petrovich}, C., {Tremaine}, S., \& {Rafikov}, R. 2014, \apj, 786, 101

\bibitem[{{Pollacco} {et~al.}(2006){Pollacco}, {Skillen}, {Collier Cameron},
  {Christian}, {Hellier}, {Irwin}, {Lister}, {Street}, {West}, {Anderson},
  {Clarkson}, {Deeg}, {Enoch}, {Evans}, {Fitzsimmons}, {Haswell}, {Hodgkin},
  {Horne}, {Kane}, {Keenan}, {Maxted}, {Norton}, {Osborne}, {Parley}, {Ryans},
  {Smalley}, {Wheatley}, \& {Wilson}}]{pollacco:2006}
{Pollacco}, D.~L., {Skillen}, I., {Collier Cameron}, A., {et~al.} 2006, \pasp,
  118, 1407

\bibitem[{{Pollack} {et~al.}(1996){Pollack}, {Hubickyj}, {Bodenheimer},
  {Lissauer}, {Podolak}, \& {Greenzweig}}]{pollack:96}
{Pollack}, J.~B., {Hubickyj}, O., {Bodenheimer}, P., {et~al.} 1996, \icarus,
  124, 62

\bibitem[{{Rabus} {et~al.}(2016){Rabus}, {Jord{\'a}n}, {Hartman}, {Bakos},
  {Espinoza}, {Brahm}, {Penev}, {Ciceri}, {Zhou}, {Bayliss}, {Mancini},
  {Bhatti}, {de Val-Borro}, {Csbury}, {Sato}, {Tan}, {Henning}, {Schmidt},
  {Bento}, {Suc}, {Noyes}, {L{\'a}z{\'a}r}, {Papp}, \& {S{\'a}ri}}]{rabus:2016}
{Rabus}, M., {Jord{\'a}n}, A., {Hartman}, J.~D., {et~al.} 2016, \aj, 152, 88

\bibitem[{{Ricker} {et~al.}(2015){Ricker}, {Winn}, {Vanderspek}, {Latham},
  {Bakos}, {Bean}, {Berta-Thompson}, {Brown}, {Buchhave}, {Butler}, {Butler},
  {Chaplin}, {Charbonneau}, {Christensen-Dalsgaard}, {Clampin}, {Deming},
  {Doty}, {De Lee}, {Dressing}, {Dunham}, {Endl}, {Fressin}, {Ge}, {Henning},
  {Holman}, {Howard}, {Ida}, {Jenkins}, {Jernigan}, {Johnson}, {Kaltenegger},
  {Kawai}, {Kjeldsen}, {Laughlin}, {Levine}, {Lin}, {Lissauer}, {MacQueen},
  {Marcy}, {McCullough}, {Morton}, {Narita}, {Paegert}, {Palle}, {Pepe},
  {Pepper}, {Quirrenbach}, {Rinehart}, {Sasselov}, {Sato}, {Seager},
  {Sozzetti}, {Stassun}, {Sullivan}, {Szentgyorgyi}, {Torres}, {Udry}, \&
  {Villasenor}}]{tess}
{Ricker}, G.~R., {Winn}, J.~N., {Vanderspek}, R., {et~al.} 2015, Journal of
  Astronomical Telescopes, Instruments, and Systems, 1, 014003

\bibitem[{{Schlaufman} \& {Winn}(2013)}]{schlaufman:2013}
{Schlaufman}, K.~C., \& {Winn}, J.~N. 2013, \apj, 772, 143

\bibitem[{{Shporer} {et~al.}(2017){Shporer}, {Zhou}, {Fulton}, {Vanderburg},
  {Espinoza}, {Collins}, {Ciardi}, {Bayliss}, {Armstrong}, {Bento}, {Bouchy},
  {Cochran}, {Collier Cameron}, {Col{\'o}n}, {Crossfield}, {Dragomir},
  {Howard}, {Howell}, {Isaacson}, {Kielkopf}, {Murgas}, {Sefako}, {Sinukoff},
  {Siverd}, \& {Udry}}]{shporer:2017}
{Shporer}, A., {Zhou}, G., {Fulton}, B.~J., {et~al.} 2017, \aj, 154, 188

\bibitem[{{Smith} {et~al.}(2013){Smith}, {Anderson}, {Bouchy}, {Collier
  Cameron}, {Doyle}, {Fumel}, {Gillon}, {H{\'e}brard}, {Hellier}, {Jehin},
  {Lendl}, {Maxted}, {Moutou}, {Pepe}, {Pollacco}, {Queloz}, {Santerne},
  {Segransan}, {Smalley}, {Southworth}, {Triaud}, {Udry}, \&
  {West}}]{smith:2013}
{Smith}, A.~M.~S., {Anderson}, D.~R., {Bouchy}, F., {et~al.} 2013, \aap, 552,
  A120

\bibitem[{{Smith} {et~al.}(2017){Smith}, {Gandolfi}, {Barrag{\'a}n}, {Bowler},
  {Csizmadia}, {Endl}, {Fridlund}, {Grziwa}, {Guenther}, {Hatzes}, {Nowak},
  {Albrecht}, {Alonso}, {Cabrera}, {Cochran}, {Deeg}, {Cusano},
  {Eigm{\"u}ller}, {Erikson}, {Hidalgo}, {Hirano}, {Johnson}, {Korth}, {Mann},
  {Narita}, {Nespral}, {Palle}, {P{\"a}tzold}, {Prieto-Arranz}, {Rauer},
  {Ribas}, {Tingley}, \& {Wolthoff}}]{smith:2017}
{Smith}, A.~M.~S., {Gandolfi}, D., {Barrag{\'a}n}, O., {et~al.} 2017, \mnras,
  464, 2708

\bibitem[{{Spake} {et~al.}(2018){Spake}, {Sing}, {Evans}, {Oklop{\v c}i{\'c}},
  {Bourrier}, {Kreidberg}, {Rackham}, {Irwin}, {Ehrenreich}, {Wyttenbach},
  {Wakeford}, {Zhou}, {Chubb}, {Nikolov}, {Goyal}, {Henry}, {Williamson},
  {Blumenthal}, {Anderson}, {Hellier}, {Charbonneau}, {Udry}, \&
  {Madhusudhan}}]{spake:2018}
{Spake}, J.~J., {Sing}, D.~K., {Evans}, T.~M., {et~al.} 2018, \nat, 557, 68

\bibitem[{{Talens} {et~al.}(2017){Talens}, {Spronck}, {Lesage}, {Otten},
  {Stuik}, {Pollacco}, \& {Snellen}}]{talens:2017}
{Talens}, G.~J.~J., {Spronck}, J.~F.~P., {Lesage}, A.-L., {et~al.} 2017, \aap,
  601, A11

\bibitem[{{Triaud} {et~al.}(2010){Triaud}, {Collier Cameron}, {Queloz},
  {Anderson}, {Gillon}, {Hebb}, {Hellier}, {Loeillet}, {Maxted}, {Mayor},
  {Pepe}, {Pollacco}, {S{\'e}gransan}, {Smalley}, {Udry}, {West}, \&
  {Wheatley}}]{triaud:2010}
{Triaud}, A.~H.~M.~J., {Collier Cameron}, A., {Queloz}, D., {et~al.} 2010,
  \aap, 524, A25

\bibitem[{{Van Eylen} {et~al.}(2016){Van Eylen}, {Albrecht}, {Gandolfi}, {Dai},
  {Winn}, {Hirano}, {Narita}, {Bruntt}, {Prieto-Arranz}, {B{\'e}jar}, {Nowak},
  {Lund}, {Palle}, {Ribas}, {Sanchis-Ojeda}, {Yu}, {Arriagada}, {Butler},
  {Crane}, {Handberg}, {Deeg}, {Jessen-Hansen}, {Johnson}, {Nespral}, {Rogers},
  {Ryu}, {Shectman}, {Shrotriya}, {Slumstrup}, {Takeda}, {Teske}, {Thompson},
  {Vanderburg}, \& {Wittenmyer}}]{vaneylen:2016}
{Van Eylen}, V., {Albrecht}, S., {Gandolfi}, D., {et~al.} 2016, \aj, 152, 143

\bibitem[{{Vanzi} {et~al.}(2018){Vanzi}, {Zapata}, {Flores}, {Brahm}, {Tala
  Pinto}, {Rukdee}, {Jones}, {Ropert}, {Shen}, {Ramirez}, {Suc}, {Jordan}, \&
  {Espinoza}}]{vanzi:2018}
{Vanzi}, L., {Zapata}, A., {Flores}, M., {et~al.} 2018, ArXiv e-prints,
  arXiv:1804.07441

\bibitem[{{Villaver} \& {Livio}(2009)}]{villaver:2009}
{Villaver}, E., \& {Livio}, M. 2009, \apjl, 705, L81

\bibitem[{{Villaver} {et~al.}(2014){Villaver}, {Livio}, {Mustill}, \&
  {Siess}}]{villaver:2014}
{Villaver}, E., {Livio}, M., {Mustill}, A.~J., \& {Siess}, L. 2014, \apj, 794,
  3

\bibitem[{{Wang} {et~al.}(2018){Wang}, {Jones}, {Shporer}, {Fulton}, {Paredes},
  {Trifonov}, {Kossakowski}, {Eastman}, {Gunther}, {Huang}, {Millholland},
  {Seligman}, {Fischer}, {Brahm}, {Wang}, {Cruz}, {James}, {Addison}, {Henry},
  {Liang}, {Davis}, {Tronsgaard}, {Worku}, {Brewer}, {Kurster}, {Beichman},
  {Bieryla}, {Brown}, {Christiansen}, {Ciardi}, {Collins}, {Esquerdo},
  {Howard}, {Isaacson}, {Latham}, {Mazeh}, {Petigura}, {Quinn}, {Shahaf},
  {Siverd}, {Ricker}, {Vanderspek}, {Seager}, {Winn}, {Jenkins}, {Boyd},
  {Furesz}, {Henze}, {Levine}, {Morris}, {Paegert}, {Stassun}, {Ting}, {Vezie},
  \& {Laughlin}}]{wang:2018}
{Wang}, S., {Jones}, M., {Shporer}, A., {et~al.} 2018, ArXiv e-prints,
  arXiv:1810.02341

\bibitem[{Welsh {et~al.}(2010)Welsh, Orosz, Seager, Fortney, Jenkins, Rowe,
  Koch, \& Borucki}]{welsh:2010}
Welsh, W.~F., Orosz, J.~A., Seager, S., {et~al.} 2010, The Astrophysical
  Journal Letters, 713, L145.
\newblock \url{http://stacks.iop.org/2041-8205/713/i=2/a=L145}

\bibitem[{{Yi} {et~al.}(2001){Yi}, {Demarque}, {Kim}, {Lee}, {Ree}, {Lejeune},
  \& {Barnes}}]{yi:2001}
{Yi}, S., {Demarque}, P., {Kim}, Y.-C., {et~al.} 2001, \apjs, 136, 417

\bibitem[{{Yu} {et~al.}(2018){Yu}, {Rodriguez}, {Eastman}, {Crossfield},
  {Shporer}, {Gaudi}, {Burt}, {Fulton}, {Sinukoff}, {Howard}, {Isaacson},
  {Kosiarek}, {Ciardi}, {Schlieder}, {Penev}, {Vanderburg}, {Stassun},
  {Bieryla}, {Butler}, {Berlind}, {Calkins}, {Esquerdo}, {Latham}, {Murawski},
  {Stevens}, {Petigura}, {Kreidberg}, \& {Bristow}}]{k2-234}
{Yu}, L., {Rodriguez}, J.~E., {Eastman}, J.~D., {et~al.} 2018, ArXiv e-prints,
  arXiv:1803.02858

\bibitem[{{Zechmeister} \& {K{\"u}rster}(2009)}]{zechmeister:2009}
{Zechmeister}, M., \& {K{\"u}rster}, M. 2009, \aap, 496, 577

\bibitem[{{Zhou} {et~al.}(2015){Zhou}, {Bayliss}, {Hartman}, {Fulton}, {Bakos},
  {Howard}, {Isaacson}, {Marcy}, {Schmidt}, {Brahm}, \&
  {Jord{\'a}n}}]{zhou:2015}
{Zhou}, G., {Bayliss}, D., {Hartman}, J.~D., {et~al.} 2015, \apjl, 814, L16

\end{thebibliography}




\appendix
\setcounter{table}{2}
\begin{longtable*}{lrrrrrrl}
\caption{Relative radial velocities and bisector spans for \stname.\label{tab:rvs}}\\
\hline
\hline
\multicolumn{1}{l}{BJD} & \multicolumn{1}{r}{RV} & \multicolumn{1}{r}{$\sigma_{\rm RV}$} & \multicolumn{1}{r}{BIS} & \multicolumn{1}{r}{$\sigma_{\rm BIS}$} & 
\multicolumn{1}{r}{S index} & \multicolumn{1}{r}{$\sigma_{\rm S index}$} & \multicolumn{1}{l}{Instrument} \\
\multicolumn{1}{l}{\hbox{(2,400,000$+$)}} & \multicolumn{1}{r}{(km s$^{-1}$)} & \multicolumn{1}{r}{(km s$^{-1}$)} & \multicolumn{1}{r}{(km s$^{-1}$)} & \multicolumn{1}{r}{(km s$^{-1}$)} & \multicolumn{1}{l}{} \\
\hline
\endfirsthead
\endhead
\endlastfoot
$ 2458367.68314629 $ & $ 30.7863 $ & $ 0.0020 $ & $  0.022 $ & $ 0.002 $ & \dotfill & \dotfill & HARPS   \\
$ 2458368.66398162 $ & $ 30.7806 $ & $ 0.0020 $ & $  0.021 $ & $ 0.002 $ & \dotfill & \dotfill & HARPS   \\
$ 2458368.68745551 $ & $ 30.6687 $ & $ 0.0066 $ & $ -0.013 $ & $ 0.006 $ & \dotfill & \dotfill & FIDEOS  \\
$ 2458368.69912355 $ & $ 30.6521 $ & $ 0.0066 $ & $ -0.028 $ & $ 0.006 $ & \dotfill & \dotfill & FIDEOS  \\
$ 2458368.70678972 $ & $ 30.6834 $ & $ 0.0073 $ & $  0.002 $ & $ 0.007 $ & \dotfill & \dotfill & FIDEOS  \\
$ 2458368.75311381 $ & $ 30.6773 $ & $ 0.0068 $ & $ -0.032 $ & $ 0.007 $ & \dotfill & \dotfill & FIDEOS  \\
$ 2458368.76041725 $ & $ 30.6538 $ & $ 0.0071 $ & $ -0.022 $ & $ 0.007 $ & \dotfill & \dotfill & FIDEOS  \\
$ 2458368.85411697 $ & $ 30.6615 $ & $ 0.0081 $ & $ -0.011 $ & $ 0.009 $ & \dotfill & \dotfill & FIDEOS  \\
$ 2458368.86145728 $ & $ 30.6624 $ & $ 0.0080 $ & $ -0.034 $ & $ 0.008 $ & \dotfill & \dotfill & FIDEOS  \\
$ 2458369.64418781 $ & $ 30.6956 $ & $ 0.0071 $ & $ -0.038 $ & $ 0.007 $ & \dotfill & \dotfill & FIDEOS  \\
$ 2458369.65198892 $ & $ 30.6926 $ & $ 0.0072 $ & $ -0.008 $ & $ 0.007 $ & \dotfill & \dotfill & FIDEOS  \\
$ 2458369.65923657 $ & $ 30.6669 $ & $ 0.0071 $ & $ -0.037 $ & $ 0.007 $ & \dotfill & \dotfill & FIDEOS  \\
$ 2458369.78679607 $ & $ 30.6421 $ & $ 0.0075 $ & $ -0.036 $ & $ 0.008 $ & \dotfill & \dotfill & FIDEOS  \\
$ 2458369.79512629 $ & $ 30.6764 $ & $ 0.0077 $ & $ -0.014 $ & $ 0.008 $ & \dotfill & \dotfill & FIDEOS  \\
$ 2458369.80270826 $ & $ 30.6555 $ & $ 0.0076 $ & $ -0.029 $ & $ 0.008 $ & \dotfill & \dotfill & FIDEOS  \\
$ 2458370.64460774 $ & $ 30.6861 $ & $ 0.0068 $ & $  0.006 $ & $ 0.007 $ & \dotfill & \dotfill & FIDEOS  \\
$ 2458370.65242178 $ & $ 30.6809 $ & $ 0.0067 $ & $  0.003 $ & $ 0.007 $ & \dotfill & \dotfill & FIDEOS  \\
$ 2458370.66037165 $ & $ 30.6822 $ & $ 0.0068 $ & $  0.005 $ & $ 0.007 $ & \dotfill & \dotfill & FIDEOS  \\
$ 2458370.78118598 $ & $ 30.6629 $ & $ 0.0079 $ & $ -0.008 $ & $ 0.008 $ & \dotfill & \dotfill & FIDEOS  \\
$ 2458370.78842912 $ & $ 30.6683 $ & $ 0.0078 $ & $ -0.016 $ & $ 0.008 $ & \dotfill & \dotfill & FIDEOS  \\
$ 2458370.79561044 $ & $ 30.6631 $ & $ 0.0079 $ & $  0.005 $ & $ 0.008 $ & \dotfill & \dotfill & FIDEOS  \\
$ 2458371.64189723 $ & $ 30.6767 $ & $ 0.0078 $ & $ -0.026 $ & $ 0.008 $ & \dotfill & \dotfill & FIDEOS  \\
$ 2458371.64982610 $ & $ 30.6960 $ & $ 0.0079 $ & $ -0.044 $ & $ 0.008 $ & \dotfill & \dotfill & FIDEOS  \\
$ 2458371.65723468 $ & $ 30.6510 $ & $ 0.0137 $ & $  0.025 $ & $ 0.014 $ & \dotfill & \dotfill & FIDEOS  \\
$ 2458371.77405182 $ & $ 30.6681 $ & $ 0.0078 $ & $ -0.005 $ & $ 0.008 $ & \dotfill & \dotfill & FIDEOS  \\
$ 2458371.78144230 $ & $ 30.6848 $ & $ 0.0078 $ & $ -0.009 $ & $ 0.008 $ & \dotfill & \dotfill & FIDEOS  \\
$ 2458371.78862670 $ & $ 30.6807 $ & $ 0.0075 $ & $ -0.019 $ & $ 0.008 $ & \dotfill & \dotfill & FIDEOS  \\
$ 2458371.87779688 $ & $ 30.6601 $ & $ 0.0095 $ & $  0.011 $ & $ 0.010 $ & \dotfill & \dotfill & FIDEOS  \\
$ 2458372.65137095 $ & $ 30.6693 $ & $ 0.0067 $ & $  0.002 $ & $ 0.007 $ & \dotfill & \dotfill & FIDEOS  \\
$ 2458372.65854816 $ & $ 30.6858 $ & $ 0.0069 $ & $ -0.018 $ & $ 0.007 $ & \dotfill & \dotfill & FIDEOS  \\
$ 2458372.67407844 $ & $ 30.6880 $ & $ 0.0067 $ & $ -0.014 $ & $ 0.007 $ & \dotfill & \dotfill & FIDEOS  \\
$ 2458372.79013200 $ & $ 30.7100 $ & $ 0.0083 $ & $  0.017 $ & $ 0.009 $ & \dotfill & \dotfill & FIDEOS  \\
$ 2458372.79735118 $ & $ 30.6948 $ & $ 0.0082 $ & $  0.005 $ & $ 0.009 $ & \dotfill & \dotfill & FIDEOS  \\
$ 2458372.80462143 $ & $ 30.6821 $ & $ 0.0083 $ & $  0.001 $ & $ 0.009 $ & \dotfill & \dotfill & FIDEOS  \\
$ 2458373.75687607 $ & $ 30.7194 $ & $ 0.0088 $ & $  0.051 $ & $ 0.009 $ & \dotfill & \dotfill & FIDEOS  \\
$ 2458373.76411946 $ & $ 30.7301 $ & $ 0.0088 $ & $  0.015 $ & $ 0.009 $ & \dotfill & \dotfill & FIDEOS  \\
$ 2458379.71952476 $ & $ 30.7427 $ & $ 0.0050 $ & $  0.001 $ & $ 0.003 $ & $ 0.1963 $ & $ 0.0019 $ & FEROS   \\
$ 2458379.72709860 $ & $ 30.7462 $ & $ 0.0050 $ & $  0.002 $ & $ 0.003 $ & $ 0.1942 $ & $ 0.0018 $ & FEROS   \\
$ 2458379.73465009 $ & $ 30.7490 $ & $ 0.0050 $ & $  0.001 $ & $ 0.003 $ & $ 0.1923 $ & $ 0.0019 $ & FEROS   \\
$ 2458383.71877730 $ & $ 30.7464 $ & $ 0.0050 $ & $  0.004 $ & $ 0.003 $ & $ 0.1905 $ & $ 0.0018 $ & FEROS   \\
$ 2458383.72286419 $ & $ 30.7444 $ & $ 0.0050 $ & $  0.002 $ & $ 0.003 $ & $ 0.1941 $ & $ 0.0018 $ & FEROS   \\
$ 2458383.72928826 $ & $ 30.7401 $ & $ 0.0050 $ & $  0.001 $ & $ 0.003 $ & $ 0.2323 $ & $ 0.0021 $ & FEROS   \\
$ 2458384.74720465 $ & $ 30.7524 $ & $ 0.0050 $ & $  0.005 $ & $ 0.003 $ & $ 0.1940 $ & $ 0.0019 $ & FEROS   \\
$ 2458384.75128807 $ & $ 30.7588 $ & $ 0.0050 $ & $  0.006 $ & $ 0.003 $ & $ 0.1947 $ & $ 0.0019 $ & FEROS   \\
$ 2458384.75537333 $ & $ 30.7585 $ & $ 0.0050 $ & $  0.002 $ & $ 0.003 $ & $ 0.1933 $ & $ 0.0020 $ & FEROS   \\
$ 2458385.69639408 $ & $ 30.7720 $ & $ 0.0050 $ & $  0.001 $ & $ 0.003 $ & $ 0.1890 $ & $ 0.0017 $ & FEROS   \\
$ 2458385.70045388 $ & $ 30.7691 $ & $ 0.0050 $ & $  0.000 $ & $ 0.003 $ & $ 0.1851 $ & $ 0.0018 $ & FEROS   \\
$ 2458385.70451623 $ & $ 30.7669 $ & $ 0.0050 $ & $  0.001 $ & $ 0.003 $ & $ 0.1935 $ & $ 0.0017 $ & FEROS   \\
$ 2458398.85664355 $ & $ 30.7997 $ & $ 0.0050 $ & $ -0.002 $ & $ 0.003 $ & $ 0.1852 $ & $ 0.0030 $ & FEROS   \\
$ 2458398.86160055 $ & $ 30.7927 $ & $ 0.0050 $ & $ -0.002 $ & $ 0.003 $ & $ 0.1991 $ & $ 0.0038 $ & FEROS   \\
$ 2458398.86646831 $ & $ 30.8011 $ & $ 0.0050 $ & $  0.006 $ & $ 0.003 $ & $ 0.1789 $ & $ 0.0037 $ & FEROS   \\
$ 2458401.67766910 $ & $ 30.7623 $ & $ 0.0050 $ & $  0.002 $ & $ 0.003 $ & $ 0.2000 $ & $ 0.0018 $ & FEROS   \\
$ 2458401.68490538 $ & $ 30.7610 $ & $ 0.0050 $ & $  0.000 $ & $ 0.003 $ & $ 0.1980 $ & $ 0.0017 $ & FEROS   \\
$ 2458401.68884631 $ & $ 30.7641 $ & $ 0.0050 $ & $  0.006 $ & $ 0.003 $ & $ 0.1973 $ & $ 0.0018 $ & FEROS   \\
$ 2458404.50796278 $ & $ 30.7509 $ & $ 0.0050 $ & $  0.006 $ & $ 0.003 $ & $ 0.1955 $ & $ 0.0028 $ & FEROS   \\
$ 2458406.54669111 $ & $ 30.7631 $ & $ 0.0050 $ & $  0.006 $ & $ 0.003 $ & $ 0.2147 $ & $ 0.0025 $ & FEROS   \\
$ 2458406.55379532 $ & $ 30.7593 $ & $ 0.0050 $ & $  0.005 $ & $ 0.003 $ & $ 0.2109 $ & $ 0.0025 $ & FEROS   \\
$ 2458406.55921117 $ & $ 30.7620 $ & $ 0.0050 $ & $  0.002 $ & $ 0.003 $ & $ 0.2151 $ & $ 0.0025 $ & FEROS   \\
$ 2458407.51309835 $ & $ 30.7715 $ & $ 0.0050 $ & $  0.008 $ & $ 0.003 $ & $ 0.2121 $ & $ 0.0026 $ & FEROS   \\
$ 2458407.51829310 $ & $ 30.7741 $ & $ 0.0050 $ & $  0.004 $ & $ 0.003 $ & $ 0.2008 $ & $ 0.0034 $ & FEROS   \\
$ 2458407.52382158 $ & $ 30.7766 $ & $ 0.0050 $ & $  0.015 $ & $ 0.003 $ & $ 0.2121 $ & $ 0.0035 $ & FEROS   \\
$ 2458409.61580549 $ & $ 30.8010 $ & $ 0.0050 $ & $  0.008 $ & $ 0.003 $ & $ 0.2164 $ & $ 0.0020 $ & FEROS   \\
$ 2458409.62127285 $ & $ 30.8011 $ & $ 0.0050 $ & $  0.003 $ & $ 0.003 $ & $ 0.2168 $ & $ 0.0020 $ & FEROS   \\
$ 2458410.69117440 $ & $ 30.8143 $ & $ 0.0050 $ & $  0.003 $ & $ 0.003 $ & $ 0.2107 $ & $ 0.0024 $ & FEROS   \\
$ 2458411.68527663 $ & $ 30.8072 $ & $ 0.0050 $ & $  0.002 $ & $ 0.003 $ & $ 0.2167 $ & $ 0.0020 $ & FEROS   \\
$ 2458411.68936375 $ & $ 30.8089 $ & $ 0.0050 $ & $  0.005 $ & $ 0.003 $ & $ 0.2136 $ & $ 0.0020 $ & FEROS   \\
$ 2458411.69344901 $ & $ 30.8056 $ & $ 0.0050 $ & $  0.009 $ & $ 0.003 $ & $ 0.2176 $ & $ 0.0021 $ & FEROS   \\
$ 2458412.65419276 $ & $ 30.7916 $ & $ 0.0050 $ & $  0.003 $ & $ 0.003 $ & $ 0.2101 $ & $ 0.0016 $ & FEROS   \\
$ 2458412.65825916 $ & $ 30.7853 $ & $ 0.0050 $ & $  0.004 $ & $ 0.003 $ & $ 0.2110 $ & $ 0.0017 $ & FEROS   \\
$ 2458412.66232846 $ & $ 30.7872 $ & $ 0.0050 $ & $  0.003 $ & $ 0.003 $ & $ 0.2101 $ & $ 0.0017 $ & FEROS   \\
$ 2458413.56187674 $ & $ 30.7734 $ & $ 0.0050 $ & $  0.005 $ & $ 0.003 $ & $ 0.2067 $ & $ 0.0015 $ & FEROS   \\
$ 2458413.56594013 $ & $ 30.7713 $ & $ 0.0050 $ & $  0.004 $ & $ 0.003 $ & $ 0.2101 $ & $ 0.0016 $ & FEROS   \\
$ 2458413.57000699 $ & $ 30.7723 $ & $ 0.0050 $ & $  0.003 $ & $ 0.003 $ & $ 0.2073 $ & $ 0.0017 $ & FEROS   \\
$ 2458414.64539175 $ & $ 30.7621 $ & $ 0.0050 $ & $  0.006 $ & $ 0.006 $ & $ 0.2081 $ & $ 0.0019 $ & FEROS   \\
$ 2458414.64946278 $ & $ 30.7646 $ & $ 0.0050 $ & $  0.007 $ & $ 0.006 $ & $ 0.2079 $ & $ 0.0017 $ & FEROS   \\
$ 2458414.65458118 $ & $ 30.7597 $ & $ 0.0050 $ & $  0.003 $ & $ 0.006 $ & $ 0.2087 $ & $ 0.0018 $ & FEROS   \\
$ 2458415.63730889 $ & $ 30.7569 $ & $ 0.0050 $ & $  0.004 $ & $ 0.007 $ & $ 0.2119 $ & $ 0.0020 $ & FEROS   \\
$ 2458415.64138930 $ & $ 30.7544 $ & $ 0.0050 $ & $  0.004 $ & $ 0.006 $ & $ 0.2046 $ & $ 0.0020 $ & FEROS   \\
$ 2458415.64547434 $ & $ 30.7619 $ & $ 0.0050 $ & $  0.005 $ & $ 0.007 $ & $ 0.2072 $ & $ 0.0021 $ & FEROS   \\
$ 2458416.60613278 $ & $ 30.7547 $ & $ 0.0050 $ & $  0.002 $ & $ 0.006 $ & $ 0.2076 $ & $ 0.0016 $ & FEROS   \\
$ 2458416.61019593 $ & $ 30.7487 $ & $ 0.0050 $ & $  0.003 $ & $ 0.006 $ & $ 0.2059 $ & $ 0.0017 $ & FEROS   \\
$ 2458416.61426082 $ & $ 30.7570 $ & $ 0.0050 $ & $  0.004 $ & $ 0.006 $ & $ 0.2040 $ & $ 0.0016 $ & FEROS   \\
$ 2458417.52542561 $ & $ 30.7916 $ & $ 0.0020 $ & $  0.023 $ & $ 0.002 $ & \dotfill & \dotfill & HARPS   \\
$ 2458418.61188680 $ & $ 30.7580 $ & $ 0.0050 $ & $  0.005 $ & $ 0.006 $ & $ 0.20121 $ & $ 0.0015541 $ & FEROS   \\
$ 2458418.61596421 $ & $ 30.7608 $ & $ 0.0050 $ & $  0.004 $ & $ 0.006 $ & $ 0.20046 $ & $ 0.0016165 $ & FEROS   \\
$ 2458418.62002979 $ & $ 30.7593 $ & $ 0.0050 $ & $  0.007 $ & $ 0.006 $ & $ 0.19964 $ & $ 0.0016737 $ & FEROS   \\
$ 2458419.58562321 $ & $ 30.7659 $ & $ 0.0050 $ & $  0.005 $ & $ 0.006 $ & $ 0.20384 $ & $ 0.0017164 $ & FEROS   \\
$ 2458419.58968614 $ & $ 30.7663 $ & $ 0.0050 $ & $  0.001 $ & $ 0.006 $ & $ 0.20433 $ & $ 0.0017265 $ & FEROS   \\
$ 2458419.58156051 $ & $ 30.7685 $ & $ 0.0050 $ & $  0.001 $ & $ 0.006 $ & $ 0.20151 $ & $ 0.0016520 $ & FEROS   \\
$ 2458419.64824552 $ & $ 30.8056 $ & $ 0.0020 $ & $  0.023 $ & $ 0.002 $ & \dotfill & \dotfill & HARPS   \\
\hline
\end{longtable*}

\end{document}